\documentclass[ reprint, twocolumn, superscriptaddress, floatfix, amsmath,amssymb, aps ]{revtex4-2}

\usepackage{graphicx}
\usepackage[english]{babel}
\usepackage{bm}
\usepackage{physics}
\usepackage{siunitx}
\usepackage[colorlinks=true,linkcolor=blue,citecolor=blue,urlcolor=blue]{hyperref}

\newcommand{\bs}[1]{\boldsymbol{#1}}
\newcommand{\qv}{\vb{q}}
\newcommand{\kv}{\vb{k}}
\newcommand{\bv}{\bs{\beta}}
\newcommand{\gt}{>}
\renewcommand{\Re}{\operatorname{Re}}

\begin{document}


\title{Many-mode grating couplers by avoiding undesired couplings}

\author{Nazar Pyvovar}
\affiliation{Department of Applied Physics and Energy Sciences Institute, Yale University, New Haven, CT, USA}
\author{Hao Li}
\affiliation{Department of ECE, Yale University, New Haven, CT, USA}
\author{Zhaowei Dai}
\affiliation{Department of Applied Physics and Energy Sciences Institute, Yale University, New Haven, CT, USA}
\affiliation{Department of ECE, Yale University, New Haven, CT, USA}
\author{Owen D. Miller}
\email[Corresponding author: ]{owen.miller@yale.edu}
\affiliation{Department of Applied Physics and Energy Sciences Institute, Yale University, New Haven, CT, USA}
\affiliation{Department of ECE, Yale University, New Haven, CT, USA}

\date{\today}

\begin{abstract}
    To couple many independent modes from free space to on chip, the key challenge is not enhancing the many necessary coupling rates (scattering-matrix elements) between targeted mode pairs. Instead, the key is to avoid additional cross-couplings to undesired modes, due to the presence of multiple simultaneously satisfied phase-matching conditions. With this principle, we identify scaling laws for the maximum number of high-efficiency multi-mode couplings that may be achievable for a given refractive index and design region, which are strongly supported by extensive numerical inverse-design experiments in 2D (one-dimensional coupler patterns, scattering in 2D). For such couplers, typical mode counts of 5--10 appear achievable. Three-dimensional couplers (patterned across two dimensions) can be markedly better, with tens of Fourier components in a single-layer device offering the possibility of high-efficiency coupling of hundreds to thousands of modes in relatively compact form factors. Numerical simulations of such a device, without any parameter optimization, predict efficiencies on the order of 5\% for 100 modes---a collective order-of-magnitude improvement over previous designs.
\end{abstract}

\maketitle

\section{Introduction}
In this paper, we design many-mode grating couplers, for coupling tens to hundreds or thousands of free-space waves to waveguide modes, and develop theoretical principles for enumerating how many independent, high-efficiency couplings should be achievable, for a given coupler area, refractive index, and number of layers. Whereas previous computations and experiments exhibit reduced efficiencies when coupling multiple modes, and especially at large mode counts~\cite{vuckovic_multifreq,fohrmann}, we show theoretically that many-mode, high-efficiency couplers should be possible. The central hurdle to overcome is undesired cross-couplings between input and output modes, which can spoil high average efficiencies. We identify these hurdles, approaches to circumvent them, and consequent multi-functional scaling laws in ``2D'' grating couplers (translation-invariant along a third dimension). The refractive index and number of designable layers are the key levers for high-efficiency, many-mode coupling. Inverse design~\cite{jensen2011topology,miller2012photonic,lu2013nanophotonic,Christiansen2021-qb} enables extensive numerical tests probing the maximum number of achievable high-efficiency couplings, whose linear scaling matches our theoretical predictions. For typical optical-frequency refractive indices, few-layer devices with high efficiencies for five or ten modes should be possible. In 3D, the situation is considerably improved: the azimuthal degree of freedom increases by an order of magnitude the number of achievable couplings. We design a single-layer, $100\lambda \times 100\lambda$ device that can couple more than 150 independent modes excited from a fiber into the fundamental TE mode (propagating in various directions) in a slab waveguide. Unable to simulate the large coupler sizes ($\sim 10,000\lambda^2$ areas for 3D structures), we design the coupler in the single-scattering limit, intentionally restricting its coupling efficiencies to 2--10\% where the single-scattering approximation is accurate. Demonstrating coupling of 150 modes at 2--10\% efficiency would already represent an order-of-magnitude improvement over current state-of-the-art \cite{fohrmann,3d_gc_group2_1,vuckovic_multifreq}, but by preventing cross-talk in these devices, we expect that they will be scalable to high efficiencies while retaining high (100+) mode count. To demonstrate that our theory is predictive beyond the single-scattering approximation, we perform full-wave simulations of a smaller $20\lambda\times20\lambda$ coupler, demonstrating coupling efficiencies above 4\% for approximately 100 modes. Many-mode grating couplers should open new capabilities for high-precision metrology~\cite{chip_metrology}, optical computing~\cite{logan_1,optical_computing_2}, photonic interconnects~\cite{vuckovic_interconnects}, telecommunications~\cite{telecommunications}, and more.

\begin{figure}[tb]
    \centering
    \includegraphics[width=\linewidth]{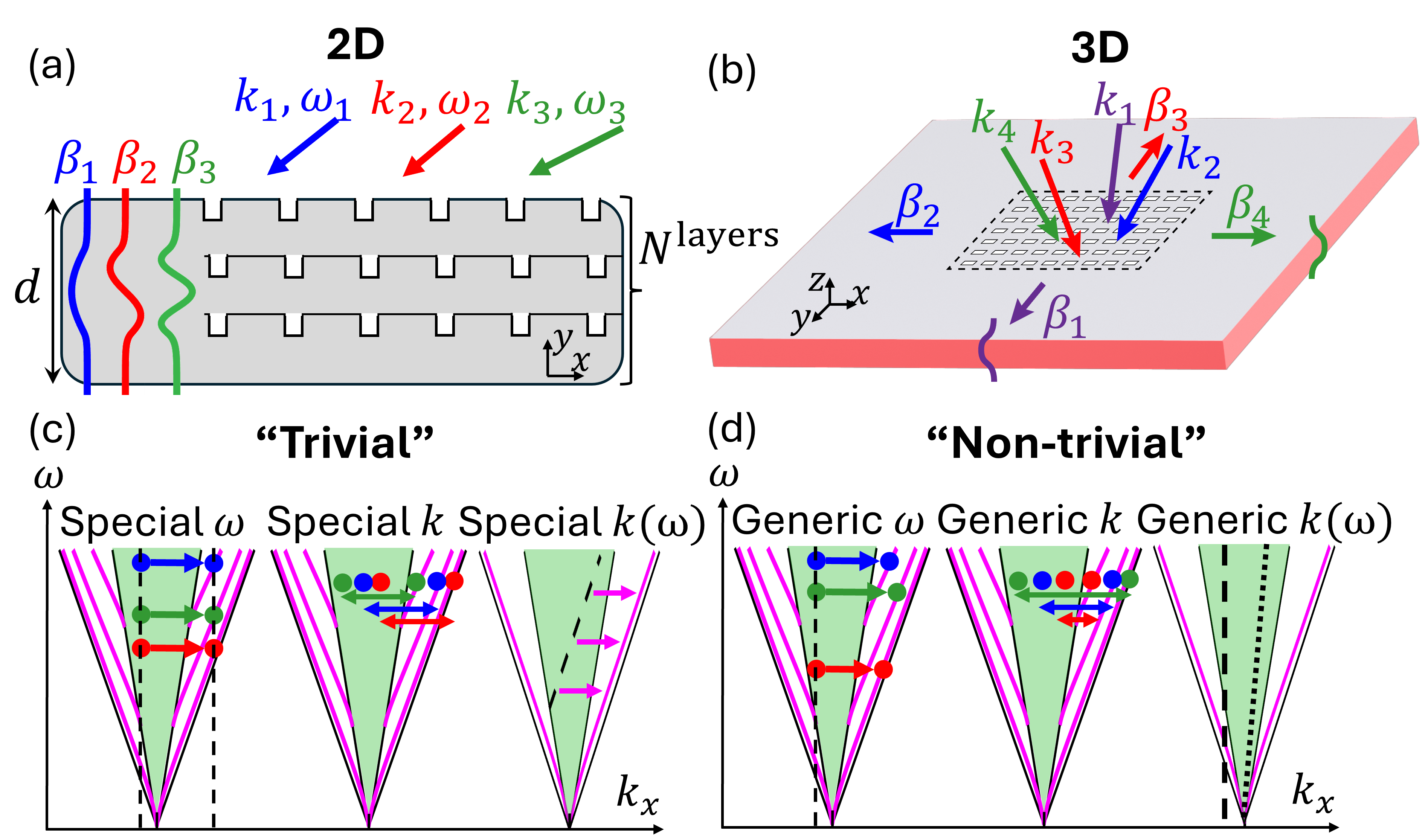}
    \caption{(a,b) Schematics of multifunctional grating couplers, coupling multiple incoming waves (launched in free space or via fiber), in (a) 2D or (b) 3D. The free-space waves have parallel wavenumbers $k_{\rm in}(\omega)$, coupled to waveguide modes with propagation constants $\beta(\omega)$. We distinguish between ``trivial'' couplings, in which special sets of input/output wavenumbers satisfy a single momentum-matching condition, as depicted in (c), versus generic, ``non-trivial'' couplings in which each target input/output pair has a different momentum-matching condition.}
    \label{fig:fig1}
\end{figure}

Grating couplers offer an interface from propagating free-space or fiber-optic waves to guided modes in integrated photonics~\cite{gc_review_2020,gc_review_2019}. Design principles of single-function grating couplers are well understood, starting from classic theory of purely periodic gratings based on phase matching and perturbation theory~\cite{early_gc_theory,theory_of_gratings,marcuse}, to more recent coupled mode theory of resonant gratings~\cite{recent_advances_in_gmr,gmr_1,gmr_2,gmr_3,resonant_gc,photonic_crystal_laser_coupled_wave}, band structure analysis~\cite{band_structure_1,band_structure_2,slanted_gc_band_structure,surface_emitting_lasers_with_blazed_outcoupling} and apodization~\cite{apod_baets1,apod_baets2,apod_fan}. Computational optimization has been used extensively to design couplers with smaller footprints~\cite{vuckovic_multifreq,hammond_johnson,replicated_opt,opt_1,opt_2,opt_3} and larger bandwidths~\cite{broadband_opt_vuckovic,broadband_opt_2,broadband_opt_3}, often guided by theoretical insights to achieve high coupling efficiencies. Overall, state-of-the-art single-function grating couplers have coupling efficiencies approaching unity~\cite{nearunity_yablonovitch,best_opt_1,bridging_the_gap}, although the last 20--30\% of efficiency may require increased fabrication complexity, including back mirrors or multi-layer designs.

Design principles of multifunctional grating couplers are significantly less explored. Computational inverse design has been used to demonstrate two-frequency~\cite{vuckovic_multifreq,vuckovic_multifreq_2,sideris_1} or two-polarization~\cite{owen_peter_gc,biwavelength_polarization_splitter} grating couplers with high efficiencies ($\gt 50\%$). A four-frequency coupler was demonstrated in~\cite{four_freq} by phase-matching the four bands of the resonant grating, creating what we call a ``trivial'' coupling scenario (cf. Fig.~\ref{fig:fig1}(c)) because a single spatial Fourier component can nominally enable all four couplings. (The term ``trivial'' is not a criticism; finding creative ways to enable applications of interest via ``trivial'' couplings may in many cases yield the greatest number of mode couplings. But trivial couplings lack flexibility in  frequency and wavevectors selection and hence are special, ``accidental'' cases, especially in 2D, i.e., for conventional devices primarily patterned across one dimension.) Designs exploiting similar trivial couplings at a single frequency, shown in Fig. 1(c) in the center, based on the analysis of the fiber mode profiles~\cite{3d_gc_group1_1,3d_gc_group1_2,3d_gc_group1_3,3d_gc_group1_4} and employing grating coupler arrays~\cite{3d_gc_group2_1,3d_gc_group2_2,3d_gc_group2_3,3d_gc_group2_4} enabled single-frequency coupling of up to ten fiber modes onto a waveguide, with experimental efficiencies of 1-15\%, well below the few-mode designs. A recent experimental work reported coupling a large subset of 2000 modes, with a reported average efficiency of 0.04\%~\cite{fohrmann}. Beyond the general lack of theoretical principles for high-efficiency, many-mode couplers, the literature also makes it appear plausible that coupling efficiencies must decrease as mode counts increase. In this paper, we aim to counteract this notion and lay out the principles for achieving high efficiencies (ultimately $\gt$ 50\%) even as mode counts scale into the hundreds, thousands, or beyond.

\section{General Theory and 2D Examples}
Design principles for single-function grating couplers originate with two essential techniques: phase-matching~\cite{theory_of_gratings} and apodization~\cite{apod_baets1}. Phase-matching prescribes that if the grating coupler is to couple an incident free-space wave peaked at wavevector $\vb k^{\text{in}}$ to a waveguide mode with wavevector $\boldsymbol{\beta}$, the grating coupler design pattern should peak at a spatial Fourier component $\vb q=\boldsymbol{\beta}-\vb k^{\text{in}}$ to ensure momentum conservation. (Throughout, every wavevector variable refers only to its component parallel to the grating surface and substrate.) Apodization refers to an additional slowly varying envelope that ensures efficient coupling to input/output beams of finite size (e.g., a Gaussian beam)~\cite{apod_baets1,apod_baets2,apod_fan}.

Consider a scenario in which one wants to couple free-space waves with (parallel) wavevectors $\vb k_i$ to waveguide modes with wavevectors $\bs\beta_i$, at frequencies $\omega_i$, with $i=1,\ldots,N_c$ for $N_c$ targeted couplings. (We consider only reciprocal systems, for which outcoupling efficiencies are identical to incoupling efficiencies.) The incoming waves interact with the scatterer's susceptibility $\chi(\omega)$, which is the difference between its permittivity and that of free space. Per phase matching, the Fourier components of the coupler's susceptibility pattern must include all \emph{differences} in the targeted free-space and waveguide wavevectors, $\vb{q}_i = \vb{k}_i - \boldsymbol{\beta}_i$:
\begin{equation}
    \chi^{\text{design}}(\bs{\rho})= \Re \sum_{i=1}^{N_c} C_i e^{i\qv_i \cdot \bs{\rho}}.
    \label{eq:chi_of_r}
\end{equation}
where $\bs{\rho}$ comprises the surface-parallel position coordinates. The first Fourier component, $\qv_1$, enables coupling $\kv_1$ to $\bs{\beta}_1$, and so on. Fourier components beyond these will also be present, to account for apodization, multiple-scattering effects, and fabrication constraints. Phase matching is the critical enabling factor for achieving moderate to high effiencies, while apodization and multiple-scattering effects take over in pushing to the efficiency limits (e.g., 80\% and beyond). Throughout, we prioritize the enabling phase-matching conditions, and leverage numerical optimization for the additional factors when feasible.

\begin{figure}[t]
    \centering
    \includegraphics[width=\linewidth]{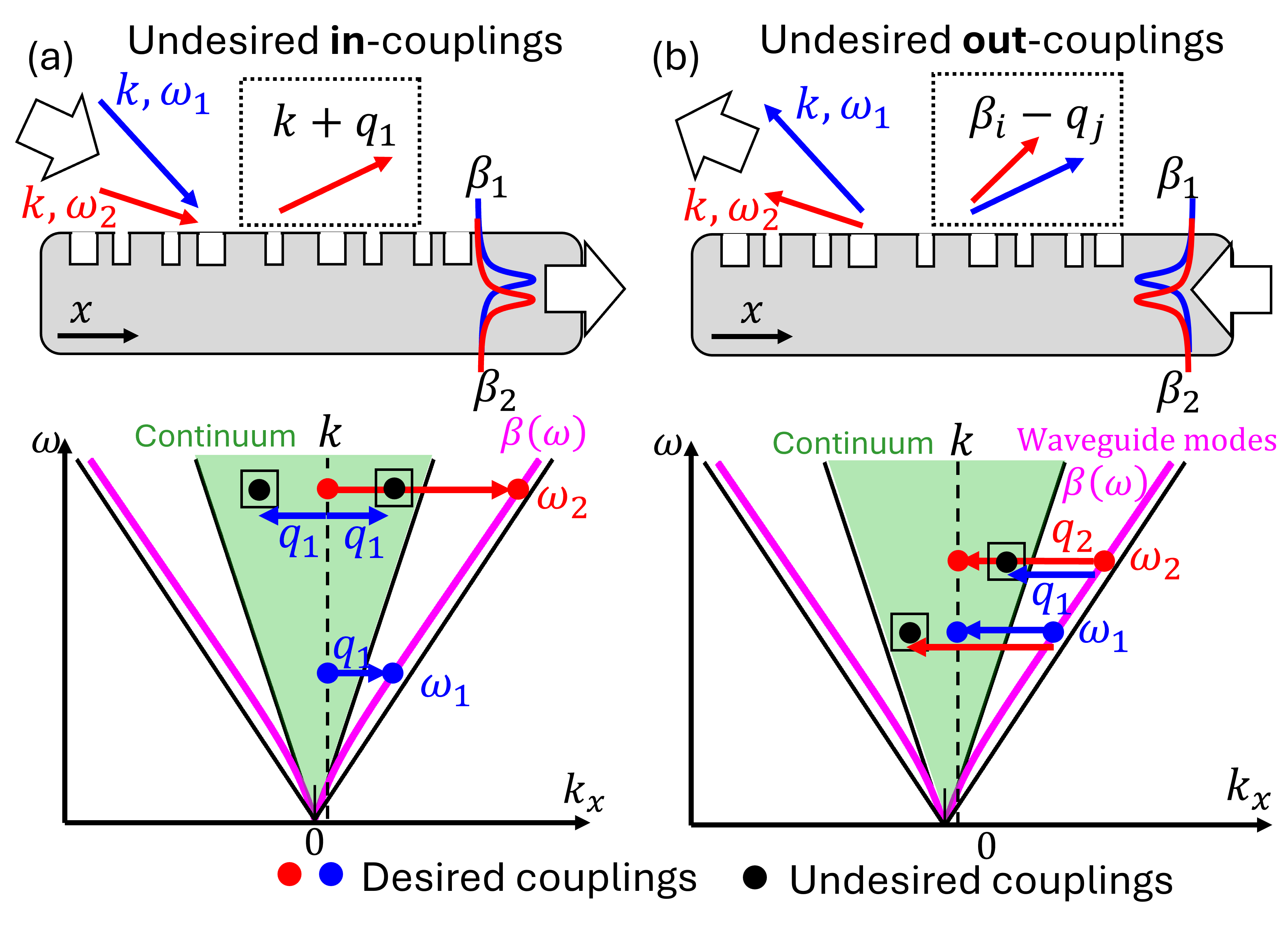}
    \caption{(a) Grating-coupler Fourier components $q_1$ and $q_2$ can create the desired couplings from incoming waves with parallel wavevector $k$ to the fundamental guided mode, at both $\omega_1$ and $\omega_2$. But they will also create undesired couplings to free-space modes at $k\pm q_1$ that make it difficult to achieve high-efficiency in-coupling. (b) Similarly, undesired \emph{out}-couplings (\emph{from} waveguide \emph{to} free-space modes) can spoil high out-coupling efficiencies; by reciprocity, they spoil in-coupling efficiencies as well. (a,b) Undesired in-couplings (out-couplings) can be avoided at smaller (larger) frequency separations, creating a fundamental tension that is difficult to circumvent.}
    \label{fig:fig2}
\end{figure}

The key challenge of multi-mode coupling is that the many spatial Fourier components of Eq.~\ref{eq:chi_of_r} not only enable the desired couplings, but also create \emph{undesired} ``cross'' couplings. Consider a two-mode coupler in which one couples $\kv_1$ to $\bv_1$ via $\qv_1$ and $\kv_2$ to $\bv_2$ via $\qv_2$. The incoming wave with $\kv_1$ will interact with $\qv_2$ (scattering to $\kv_1 \pm \qv_2$), and similarly $\qv_1$ creates couplings to $\kv_2 \pm \qv_1$. Fig.~\ref{fig:fig2}(a) shows a scenario in which the latter couplings, $\kv_2 \pm \qv_1$, occur to propagating modes in the light cone (``continuum,'' green). If the new wavevenumbers $\kv_2 \pm \qv_1$ correspond to propagating modes, a substantial fraction of incoming energy can be redirected to them (e.g., via diffractive reflection), with a corresponding loss of efficiency to the desired mode.

Perhaps surprisingly, the reciprocal scenario can introduce additional undesired couplings. Consider the reciprocal to the scenario just described, now with waveguide mode $\bs{\beta}$ launched and coupled to free-space wave $\kv$. The presence of multiple spatial Fourier coefficents can create undesired transitions from the waveguide mode to other free-space waves, as shown in Fig.~\ref{fig:fig2}(b). (There is a nomenclature subtlety in this discussion: reciprocal scenarios for incoming and outgoing ports require use of their time-reversed counterparts, e.g. $-\bs{\beta}$ and $-\kv$ for the excitations in Fig.~\ref{fig:fig2}(b). But we do not want to flip the dispersion diagram and hence do not include the negative signs in our labeling, retaining throughout the incoming-from-free-space convention. Alternatively, one can simply imagine inverting the in-plane axs directions.)  If the outgoing coupling efficiency is reduced by these undesired transitions, then, by reciprocity, the incoming efficiency equivalently suffers. Without mechanisms to control outgoing power for the undesired couplings (and we find that even every-pixel-degree-of-freedom inverse design struggles without the approaches mentioned below), the incident power will generically distribute across the desired and undesired channels equally, reducing the efficiency of the device. We propose three approaches for avoiding or inhibiting the undesired couplings.

\emph{Approach 1}: leverage ``trivial'' functionalities, such as those shown in Fig.~\ref{fig:fig1}(c), which only require a single Fourier component. As discussed above, this has been utilized in 2D in~\cite{four_freq}. In 2D, this approach imposes tight restrictions on the allowed incident modes and frequencies but, as we show below, in 3D, trivial couplings arise naturally in many coupling scenarios of interest, which should enable high-efficiency coupling of many modes.

\begin{figure}[tb]
    \centering
    \includegraphics[width=\linewidth]{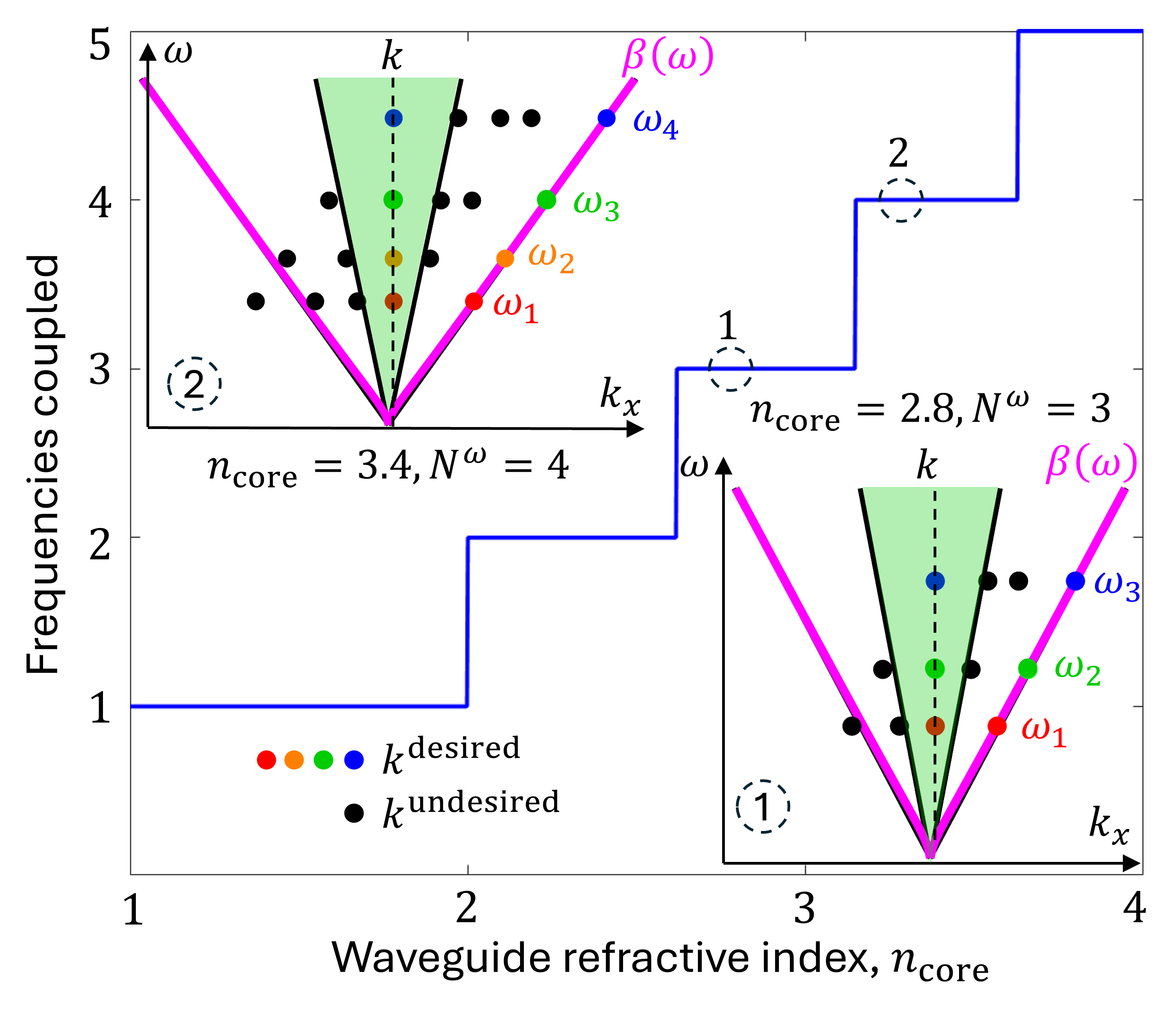}
    \caption{Maximum number of frequencies that can be coupled while avoiding undesired transitions by a single-layer grating coupler, as a function of waveguide refractive index $n_\text{core}$. Larger refractive index increases the slope of the waveguide modes, increasing the separation between necessary Fourier components and pushing undesired transitions outside the light cone, as depicted in the insets.}
    \label{fig:fig3}
\end{figure}
\begin{figure*}[tb]
    \centering
    \includegraphics[width=\linewidth]{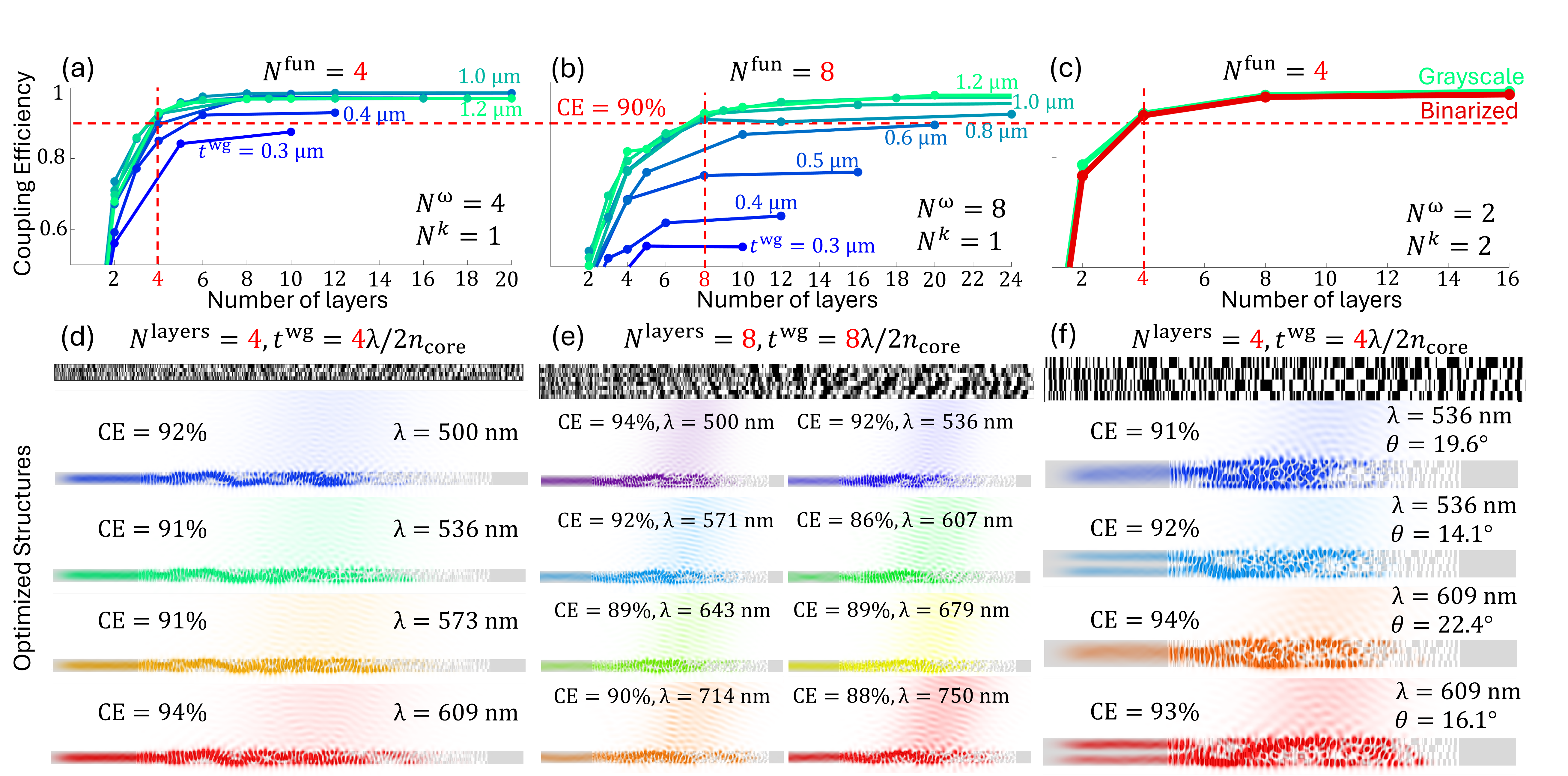}
    \caption{Computational inverse design of (a) 4-frequency, (b) 8-frequency, and (c) 2-frequency, 2-wavevector multifunctional grating couplers, coupling free-space Gaussian beams to modes of a SiN slab waveguide. Designs are divided into layers, with translational invariance in the vertical direction imposed within each layer. Varying the number of layers and total design thickness reveals scaling laws (red coefficients) consistent with our theoretical predictions. Optimized designs and their field patterns are shown for each coupling scenario in (d), (e), and (f), respectively. Additional binarization constraints were imposed in (c,f), showing little performance penalty compared to the grayscale designs.}
    \label{fig:fig4}
\end{figure*}

\emph{Approach 2}: ensure that the undesired couplings are pushed outside the free-space radiation continuum (green cone in Fig.~\ref{fig:fig1}, Fig.~\ref{fig:fig2}, Fig.~\ref{fig:fig3}), so that they do not contribute to energy loss. (``Accidental'' couplings outside the continuum to waveguide modes are sufficiently rare that they can be ignored.) This depends primarily on the separation between the coupling frequencies, as shown in the inset of Fig.~\ref{fig:fig3}. Larger frequency separations between two frequencies $\omega_1$ and $\omega_2$ can prohibit undesired couplings at $\omega_1$, though too large of a separation will create undesired couplings at $\omega_2$. There is therefore, given a frequency $\omega_1$, a frequency range $[\omega_2^\text{min},\omega_2^\text{max}]$, where another frequency $\omega_2$ may lie to avoid undesired couplings. There is also therefore a corresponding maximal set of frequencies $\omega_1,...,\omega_n$ such that each pair of frequencies $\omega_i,\omega_j$ lay within the allowed frequency ranges of each other.

The size of this maximal set can be uniquely determined, given a starting frequency $\omega_1$, by solving the inequalities that express the conditions for undesired couplings to stay outside the continuum, as equalities in their limiting cases. Two examples of such sets are shown in the insets of Fig.~\ref{fig:fig3}. The main parameter controlling the maximum number of frequencies that can be coupled this way is the refractive index $n_\text{core}$, which determines the ``separation dispersion'' between the waveguide modes and the continuum modes in the band diagram. The greater the refractive index, the greater the separation dispersion and the more room there is for pushing undesired couplings outside the continuum. By performing an exhaustive sweep over $\omega_1$ and $k$, and by finding the maximal set $\omega_1,...,\omega_n$ for each pair $(\omega_1,k)$, a maximum number of frequencies for a given core refractive index, $n_\text{core}$, can be found. Further details of this calculation may be found in the SM. As shown in Fig.~\ref{fig:fig3}, for realistic materials with refractive indices up to $n_\text{core}\leq 4$, only up to 5 frequencies may be coupled using this approach. Combining approaches 1 and 2 may allow one to couple more modes, though then the possible frequencies and wavevectors may be further restricted to satisfy the requirements of approach 1. To enable more functions or more frequency/wavenumber freedom, multilayer designs appear necessary.

\emph{Approach 3}: use multilayer designs to force undesired coupling rates near zero through destructive interference effects. It is already understood that a relative shift between two layers can cancel unwanted downward radiation~\cite{nearunity_yablonovitch}. More generally, multiple layers with appropriately shifted Fourier components can cancel the larger set of unwanted couplings that arise in many-mode couplers. Suppose each layer of a multilayer design has a susceptibility pattern following Eq.~\ref{eq:chi_of_r}, with Fourier components $\vb q_i$ common to all layers but the coefficients $C_i$ allowed to vary from layer to layer:
\begin{equation}
    \chi^{\text{layer}}(\bs{\rho})= \Re \sum_i^{N^\text{fun}} C_i^{\text{layer}}e^{i\vb q_i \cdot \bs{\rho}},
    \label{eq:chi_of_r_multilayer}
\end{equation}
such that the total number of (complex-valued) designable degrees of freedom in a multilayer design is:
\begin{equation}
    N^{\text{dof}}=N^\text{layers}\cdot N^{\vb q}=N^\text{layers}\cdot N^\text{\text{fun}},
    \label{eq:N_dof}
\end{equation}
where $N^\text{fun}$ is the number of target functions, which for generic non-trivial couplings equals the number of Fourier components $N^{\vb q}$ (cf. SM).

These designable degrees of freedom must be used to tune the undesired coupling rates to zero while keeping desired coupling rates high, resulting generically in a number of constraints equal to double the squared number of functions (cf. SM):
\begin{equation}
    N^{\text{constraints}}=2(N^\text{fun})^2.
    \label{eq:N_constraints}
\end{equation}
From Eqs.~\ref{eq:N_dof}-\ref{eq:N_constraints}, we conclude that performing $N^\text{fun}$ target functions requires at least twice as many layers:
\begin{equation}
    N^\text{layers}\geq 2N^\text{fun}.
    \label{eq:layers_scaling}
\end{equation}

We complement our theory with computational inverse design of multifunctional couplers coupling multiple free-space Gaussian beams to modes of a slab waveguide in 2D, as shown in Fig.~\ref{fig:fig4} (translational invariance assumed in the transverse direction). We consider three coupling scenarios: (1) coupling four input beams at four different frequencies, Fig.~\ref{fig:fig4} (a); (2) eight beams at eight frequencies, Fig.~\ref{fig:fig4} (c); and (3) beams at two different frequencies with two different wavevectors, with a total of four beams coupled, Fig.~\ref{fig:fig4} (e). For each scenario, we introduce a designable region where the permittivity is allowed to vary continuously between the permittivity of free-space $\epsilon^\text{free}=1$ and that of silicon nitride $\epsilon^{\operatorname{SiN}}=4$, with unpatterned $\operatorname{SiN}$ slab waveguide regions adjacent to the left and right of the design region, and implement adjoint-based topology optimization~\cite{jensen2011topology,miller2012photonic,lu2013nanophotonic,Christiansen2021-qb} to maximize the power coupled into the fundamental mode of the waveguide. We also subdivide the design region into ``layers'', with translational invariance along the vertical direction imposed within each layer. Electromagnetic simulations are performed using a volume integral equation method accelerated by a fast direct solver~\cite{Xue2023-ut}. Gradients are computed via the adjoint method, and the structure is optimized using the method of moving asymptotes~\cite{svanberg1987method,NLopt}. The design variables are initialized with random values across the design region, and each optimization is run for 1000 iterations until convergence. Each gradient-based optimization was restarted 10 times with different initial random seeds, with low-quality optima discarded. Codes reproducing the final designs shown in Fig.~\ref{fig:fig4} can be found in a Github repository listed at the end of the manuscript~\cite{github_repo}.

By varying the thicknesses and numbers of designable layers of the coupler, we find two scaling laws. First, 
the thickness of designs achieving above $90\%$ efficiency scales linearly with the number of functions, with the minimum thickness given by $t^\text{wg}\geq N^\text{fun}\frac{\lambda}{2n^\text{core}}$. Intuitively, each increment of approximately $\frac{\lambda}{2n^\text{core}}$ in thickness effectively enables one additional function. Similarly, the minimum number of layers required to surpass 90\% coupling efficiency appears to scale exactly with the number of target functions $N^\text{layers}\geq N^\text{fun}$, which agrees within a factor of 2 with our prediction of Eq.~\ref{eq:layers_scaling}. We associate this additional factor of 2 improvement with the improved unidirectionality of non-perturbative designs, whereby using only $N^\text{fun}$ layers the non-perturbative couplers are able to both perfectly cancel downward radiation and at the same time control upward coupling rates. This is a known effect in the theory of resonant gratings, where perfect unidirectional coupling may not require two parameters, as predicted by Eq.~\ref{eq:layers_scaling}, and instead can be achieved by varying a single parameter of the grating~\cite{unidirectional_soljacic}. To our knowledge, the designs in Fig.~\ref{fig:fig4} exhibit the largest number of independent couplings for high-efficiency gratings to date. And yet, the notion of using 8 patterned layers for 8 high-efficiency couplings is likely untenable. In the next section, we show that 3D potentially unlocks coupling many more modes, even with single-layer patterning, without introducing undesired couplings.

\begin{figure*}[tb]
    \centering
    \includegraphics[width=\linewidth]{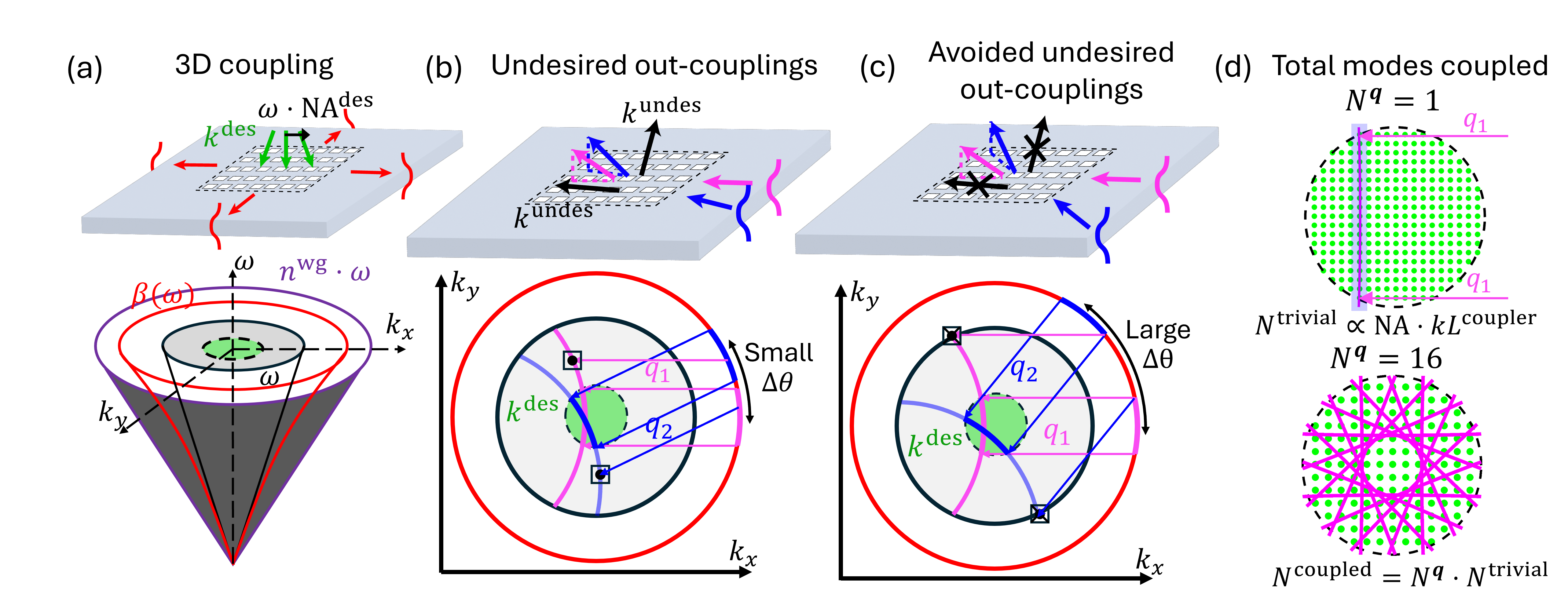}
    \caption{(a) Band diagram of the 3D coupling scenario. Desired free-space modes are contained within some fixed numerical aperture. (b) Undesired out-couplings in the 3D case occur when the angle between successive Fourier components $\vb q_1,\vb q_2$ is too small; (c) increasing the angle between $\vb q_1,\vb q_2$ ensures the undesired couplings lay outside the continuum, similar to considerations of Fig.~\ref{fig:fig3} in the 2D case. (d) A single Fourier component $\vb q_1$ may be used to couple a line of desired free-space modes. For realistic refractive indices and numerical apertures, up to 16 Fourier components can be introduced while ensuring no undesired couplings, proportionally increasing the number of coupled modes.}
    \label{fig:fig5}
\end{figure*}
\section{Many-mode 3D Couplers}
In 3D, many more modes can be coupled with a single-layer grating coupler than in 2D. The reason may be seen from Fig.~\ref{fig:fig5}, where we consider a common scenario of coupling free-space plane waves (both $s$- and $p$-polarized), within some fixed numerical aperture (NA), to the fundamental TE modes propagating in any direction within a slab waveguide. As shown in Fig.~\ref{fig:fig5}(d), even a design with a single Fourier component (shown in pink) may create many desired couplings (shown in light blue), as opposed to the 2D equivalent, which results in only a single coupling. This phenomenon is analogous to the ``trivial'' couplings discussed in the 2D case above (see Fig.~\ref{fig:fig1}), whereby a single Fourier component enables multiple equivalent-phase-matching couplings. Exploiting trivial couplings is highly beneficial for coupling to many modes. One can then superimpose the avoided-cross-couplings principle to maximize the total mode coupling count for a single-layer design.

\begin{figure*}
    \centering
    \includegraphics[width=\linewidth]{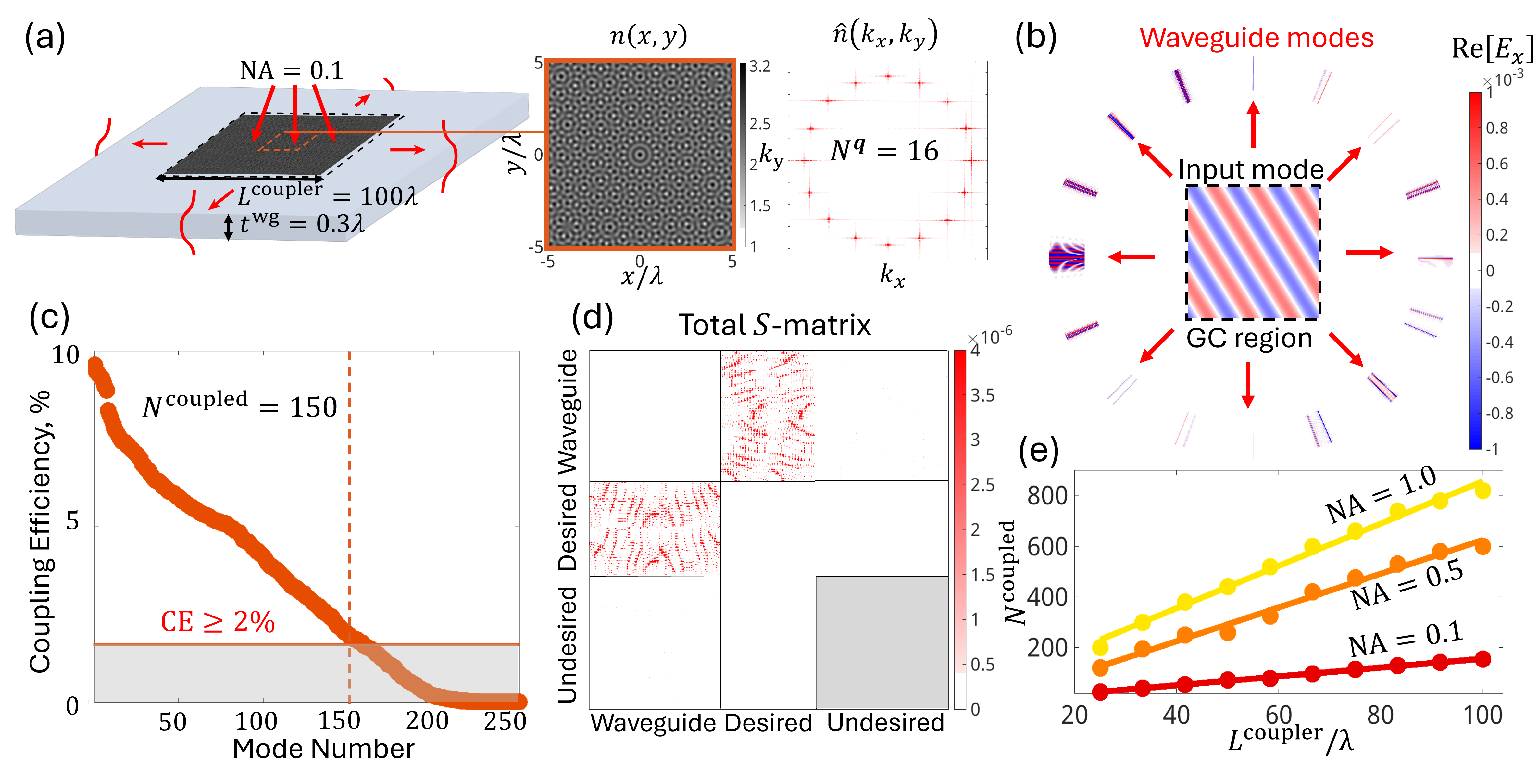}
    \caption{(a) A large ($100\lambda\times100\lambda$) perturbatively optimized design with maximal number of Fourier components, see Eq.~\ref{eq:N_Fourier}, for $n_\text{wg}=3.24$. (b) A plane wave incident on the grating coupler region is coupled to slab waveguide modes propagating along 16 directions, corresponding to the 16 Fourier components comprising the design.  (c) The design couples 150 modes with coupling efficiency uniformly above 2\%, (d) where only couplings between waveguide modes and desired free-space modes are non-zero, with all undesired couplings suppressed. (e) Maximum number of modes coupled, by rank maximization of the $S$-matrix in Eq.~\ref{eq:S_perturbative}, grows linearly with design size and desired numerical aperture, in agreement with Eq.~\ref{eq:N_total}.}
    \label{fig:fig6}
\end{figure*}

To estimate mode-coupling counts in 3D, by analogy with the 2D theory, consider a design permittivity of the form Eq.~\ref{eq:chi_of_r}, where the Fourier wavevectors $\vb q_n$ now have two in-plane components. As shown in Fig.~\ref{fig:fig5} (b), the presence of two Fourier components $\vb q_1,\vb q_2$ may result in undesired couplings, similar to the 2D case. In 3D, this happens if the angle $\Delta\theta$ between the two Fourier components is too small. As shown in Fig.~\ref{fig:fig5}(c), we find that increasing the angular separation between coupled in-plane modes can push undesired couplings outside the continuum. The minimum angular separation $\Delta\theta^\text{min}$ is determined by the effective refractive index of the waveguide modes $n^\text{eff}_\text{wg}$ and the numerical aperture of the free-space modes $\operatorname{NA}$, for which we derive a simple expression (cf. SM). The maximum possible number of Fourier components is then the number of multiples of $\Delta\theta^\text{min}$ that can fit within $2\pi$, given by:
\begin{equation}
    N^{\vb q}\leq 2\left\lfloor \frac{\pi}{\sin^{-1}\frac{1}{n^\text{eff}_\text{wg}}+\sin^{-1}\frac{\operatorname{NA}}{n^\text{eff}_\text{wg}}}\right\rfloor,
    \label{eq:N_Fourier}
\end{equation}
where the bracketed operator denotes the integer floor. For realistic values of $\operatorname{NA}=0.1$ and $n_{\text{wg}}^{\text{eff}}=3.1$, up to $N^{\vb q}=16$ may be introduced. Each Fourier component allows the ``trivial'' coupling of a line of $N^\text{trivial}\propto \operatorname{NA}\cdot (kL^\text{coupler})$ free-space modes, as shown in Fig.~\ref{fig:fig5}(d). The total number of modes coupled by $N^{\vb q}$ Fourier components is then:
\begin{equation}
    N^\text{total}_\text{coupled}\propto N^{\vb q}\cdot N^{\text{trivial}}\propto N^{\vb q}\cdot\operatorname{NA}\cdot (kL^{\text{coupler}}).
    \label{eq:N_total}
\end{equation}
Note that $N^{\vb q}$ is in turn determined by $\operatorname{NA}$ and $n_\text{wg}^\text{eff}$ via Eq.~\ref{eq:N_Fourier}.

We will now show that a $100\lambda\times 100\lambda$ design should enable coupling over $N^\text{total}_\text{coupled}>150$ modes, even for NA of only 0.1. Computational inverse design on 3D structures with such lateral dimensions, for hundreds of target incident modes, is a formidable computational task beyond current photonic simulation tools. Instead, for a proof-of-principle demonstration, we develop perturbative single-scattering designs with coupling efficiencies of 2--10\%. We then demonstrate by full-wave simulation of a smaller $20\lambda\times20\lambda$ coupler that, by increasing the strength of the perturbation (e.g., increasing the fill fraction and size of the coupler), our designs can be made high-efficiency while preserving high mode counts, owing to the avoidance of undesired transitions.

For first-order scattering processes, the $S$-matrix between the free-space and waveguide modes is given by simple mode overlaps:
\begin{equation}
    S=(M^\text{wg})^\dag \chi M^\text{free},
    \label{eq:S_perturbative}
\end{equation}
where $M^\text{wg}$ and $M^\text{free}$ are the matrices whose columns contain the discretized spatial distributions of waveguide and free-space modes respectively, and where $\chi$ is the discretized diagonal matrix of the grating coupler design. Because the expressions for waveguide and free-space modes are available analytically, the S-matrix of Eq.~\ref{eq:S_perturbative} can be evaluated very rapidly.

To obtain a design coupling $N_\text{coupled}^\text{total}>150$ modes, we numerically maximize the rank of the perturbative $S$-matrix given by Eq.~\ref{eq:S_perturbative}. To do this, we parametrize the design susceptibility $\chi^\text{design}(\vb r)$ according to Eq.~\ref{eq:chi_of_r}, where we introduce 16 Fourier components according to Eq.~\ref{eq:N_Fourier}, and where we fix the directions of the Fourier components $\hat q_i=\frac{\vb q_i}{|\vb q_i|}$ to ensure that all undesired couplings lay outside the continuum. We then optimize the remaining parameters, all $C_n$ and $|\vb q_n|$, to maximize the numerical rank of the $S$-matrix Eq.~\ref{eq:S_perturbative}. To perform the optimization, we maximize the following figure of merit:
\begin{equation}
    \operatorname{FOM}(S)=\operatorname{Tr}[(S^\dag S+\alpha)^{-1}],
\end{equation}
where ``$\operatorname{Tr}$'' denotes the matrix trace and $\alpha$ is a Tikhonoff regularization parameter. The objective originated in a computational-imaging study~\cite{kienesberger2026inprep} and is designed to increase the minimal singular value of the matrix $S$ (up to a noise floor set by $\alpha$), thus maximizing its (numerical) rank. Notice that because $S$ only contains the free-space-to-waveguide block of the full S-matrix, $S$ need not be unitary.

The results of the optimization for $L^\text{coupler}=100\lambda$ are shown in Fig.~\ref{fig:fig6} (a-d). The resulting design couples 150 free-space modes with coupling efficiencies of 2--10\%. Furthermore, as shown in Fig.~\ref{fig:fig6}(d), no undesired couplings are introduced in the total $S$-matrix of the device, which ensures that making the design larger will increase coupling efficiency. As shown in Fig.~\ref{fig:fig6}(e), similar optimizations for different values of $\operatorname{NA}$ and $L^\text{coupler}$ verify our prediction of Eq.~\ref{eq:N_total}, showing that more modes can also be coupled with larger coupler sizes and larger numerical apertures of the desired modes.

\begin{figure*}[tb]
    \centering
    \includegraphics[width=\linewidth]{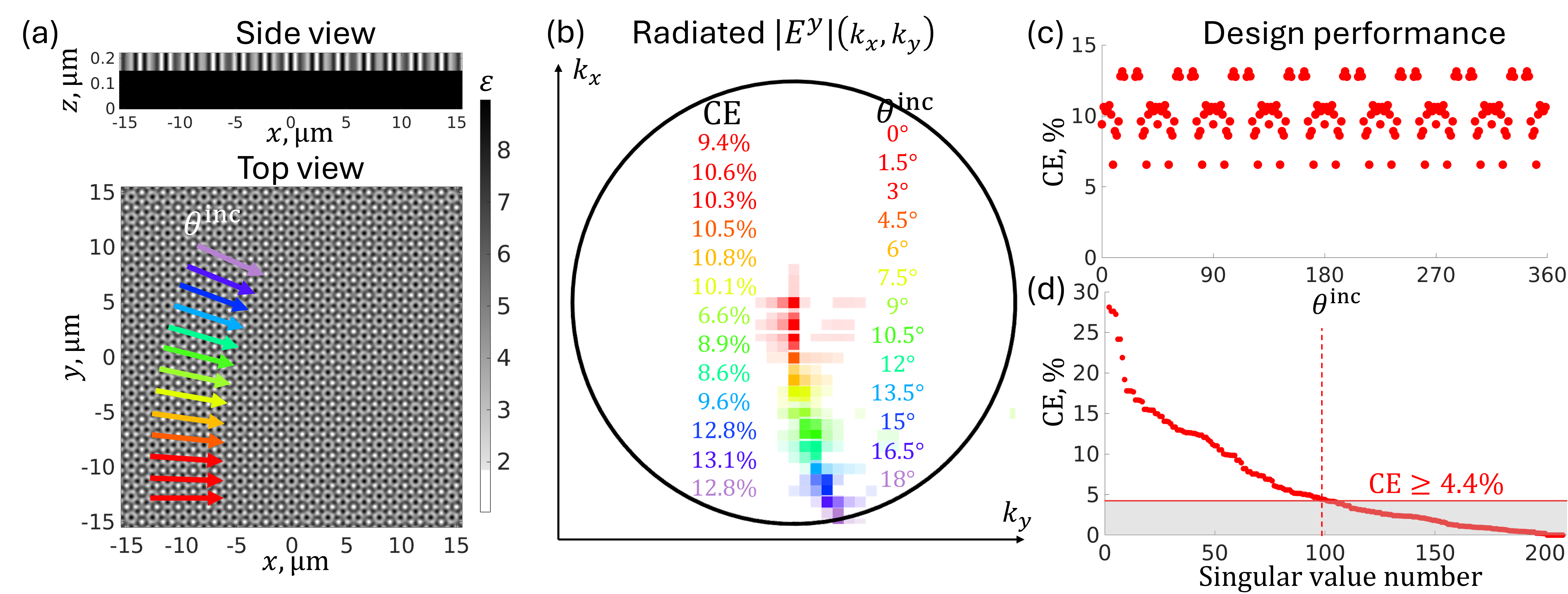}
    \caption{Full-wave simulation of a $20\lambda\times20\lambda$ grating coupler for $\lambda=\SI{1550}{nm}$. (a) The coupler has 8-fold rotational symmetry and is obtained by overlaying four periodic gratings, rotated by angles $0,\frac{\pi}{4},\frac{\pi}{2},\frac{3\pi}{4}$. Each grating has period $\Lambda=\frac{2\pi}{\beta}$, where $\beta$ is the propagation constant of the fundamental TE mode in the slab. The periods are chosen according to phase-matching, and the angles of rotation are chosen to avoid undesired couplings. (b) Slab waveguide modes incident from different angles are out-coupled to free-space modes that, in $k$-space, lay on the curves predicted by phase-matching and seen in Fig.~\ref{fig:fig5}. (c) The modes at the shown incident angles have no undesired couplings, and outcouple at efficiencies between $6.6\%$ and $13.1\%$. (d) Distribution of the singular values of the desired sector of the $S$-matrix, demonstrating, for numerical aperture 1, efficiencies between 4--30\% for more than 100 orthogonal pairs of mode couplings.}
    \label{fig:fig7}
\end{figure*}

As a complementary demonstration, in Fig.~\ref{fig:fig7} we simulate a $20\lambda\times20\lambda$ design using Flexcompute's high-performance photonic simulation tool Tidy3D~\cite{tidy3d}. While Tidy3D's extensive hardware and software optimization allows for simulating structures of this size, inverse design for multimode objectives is still out of reach due to the prohibitive cost of the many simulations required by the optimization. However, the predictions of our theory can be tested even with non-optimized, non-apodized, ``simple'' gratings. Standard single-function grating coupler theory~\cite{apod_baets1,apod_baets2,apod_fan} dictates that a slab waveguide mode incident on the grating coupler will out-couple to the phase-matched free-space mode, with total out-coupled power growing with the coupler length according to the coupling rate per unit length, which is determined by the coupler and incident field profiles~\cite{apod_fan}. This picture extends to multifunctional grating couplers if undesired couplings are avoided. Indeed, owing to the absence of undesired couplings, a slab waveguide mode incident on the coupler is only phase-matched to desired free-space modes. Therefore, increasing the size of the coupler will result in more power scattered into desired modes, yielding a device with higher coupling efficiency. This strategy may allow to achieve coupling efficiencies of up to $40\%$ for coupling to Gaussian beams, beyond which point apodization is necessary to ensure the amplitude profile of the outgoing field is Gaussian. In the multifunctional setting, where any given slab mode need not couple to a single Gaussian beam, but can instead couple to any linear combination of the desired modes, we expect the maximum coupling efficiencies that can be achieved using this strategy may be even higher.

To verify the picture described above, we simulate a coupler obtained by overlaying 4 identical periodic gratings rotated by angles of $0,\frac{\pi}{4},\frac{\pi}{2},\frac{3\pi}{4}$, as shown in Fig.~\ref{fig:fig7}(a). The resulting permittivity profile has 8-fold discrete rotational symmetry and is normalized such that the refractive index everywhere lies between $n^\text{air}=1$ and $n^\text{max}=3.0$. For further details on the design profile, cf. SM and our Github repo linked below~\cite{github_repo}. By exciting slab waveguide modes propagating along different in-waveguide-plane directions and calculating their scattered fields, we verify that the scattered fields result from phase-matching, as shown in Fig.~\ref{fig:fig7}(b). We also verify that increasing the design size results in higher coupling efficiencies, since the propagating slab waveguide modes have more length to radiate out. For the design shown in~\ref{fig:fig7}, efficiencies between $6.6\%$ and $13.1\%$ were achieved for each slab mode incident at the angles shown in~\ref{fig:fig7}(c). Calculating the total scattering matrix and taking its singular value decomposition, we find that the coupler enables 100 linearly independent couplings with efficiencies above 4\%, as shown in~\ref{fig:fig7}(d), which exceeds the existing designs in the literature both in the mode count and the average coupling efficiency \cite{fohrmann,3d_gc_group2_1,3d_gc_group2_2,3d_gc_group2_3,3d_gc_group2_4}. Further increasing the coupler size would result in both higher coupling efficiencies and higher mode counts, but doing so requires larger-scale simulation capabilities.

\section{Summary and Discussion}
In this paper, we have developed basic design principles for many-mode (or ``multi-functional'') grating couplers. In generalizing the single-mode phase-matching condition to many-mode scenarios, it is easy to identify the various Fourier components necessary to enable the desired, target ``couplings'' (scattering-matrix elements). But there are a variety of unwanted couplings that can materialize, including some non-obvious reciprocal couplings, with potentially harmful effects on attainable coupling efficiencies. The best many-mode grating coupler for scenarios of interest can benefit from a combination of three approaches: creating as many ``trivial'' couplings as the scenario allows, choosing non-trivial couplings that ensure that undesired couplings lay outside the free-space continuum, and using multilayer designs to control, through interference, those undesired couplings which cannot be avoided. Using this framework, we develop scaling laws that appear to work well even for computationally inverse-designed grating couplers, and we design numerical experiments demonstrating non-trivial coupling for hundreds of independent modes.

Looking forward, our approach offers a clear path towards high-efficiency couplers from free space to on chip, at the level of hundreds or thousands (or even more) modes. Another paper from many of the same authors~\cite{Li2026-inprep} includes a finding that enforcement of fabrication tolerances, minimum feature sizes, and robustness considerations in the computational design process only leads to modest efficiency penalties and no change in scaling laws. In 3D, beyond our discussion above, there are  additional ``knobs'' offering further degrees of freedom: leveraging multiple frequencies, accessing higher-order waveguide modes, tweaking beam-shaping coefficients, and implementing multiple designable layers. A key enabler will be effective numerical simulation at large scale, whether through adiabatic~\cite{Johnson2002-yo,Fisher2022-qr}, $\mathbb{T}$-matrix~\cite{Zhan2019-ox,Theobald2021-gp}, GPU-based~\cite{tidy3d,Egel2017-ul}, or integral-equation~\cite{Pissoort2007-qa,Xue2023-ut} approaches. With such computational tools, and in tandem with experimental demonstrations, we expect rapid progress on highly efficient, many-mode, free-space-to-on-chip couplers in the near future.

\begin{acknowledgments}
We thank many colleagues for helpful conversations, including the groups of Logan Wright and A. Douglas Stone at Yale. This work was partially supported by AFOSR grant no. FA9550-24-1-0193 and by the Simons Collaboration on Extreme Wave Phenomena Based on Symmetries (award no. SFI-MPS-EWP-00008530-09).
\end{acknowledgments}

\bibliography{refs}

\end{document}


\title{Supplementary information: \\Many-mode grating couplers by avoiding undesired couplings}

\author{Nazar Pyvovar}
\affiliation{Department of Applied Physics, Yale University, New Haven, CT, USA}

\author{Hao Li}
\affiliation{Department of Applied Physics, Yale University, New Haven, CT, USA}

\author{Zhaowei Dai}
\affiliation{Department of Applied Physics, Yale University, New Haven, CT, USA}

\author{Owen D. Miller}
\affiliation{Department of Applied Physics, Yale University, New Haven, CT, USA}

\maketitle

\tableofcontents
\section{Supplementary Information to Section II}

\subsection{Limits to the multifunctionality of single-layer 2D couplers}

\begin{figure}[h]
    \centering
    \includegraphics[width=0.75\linewidth]{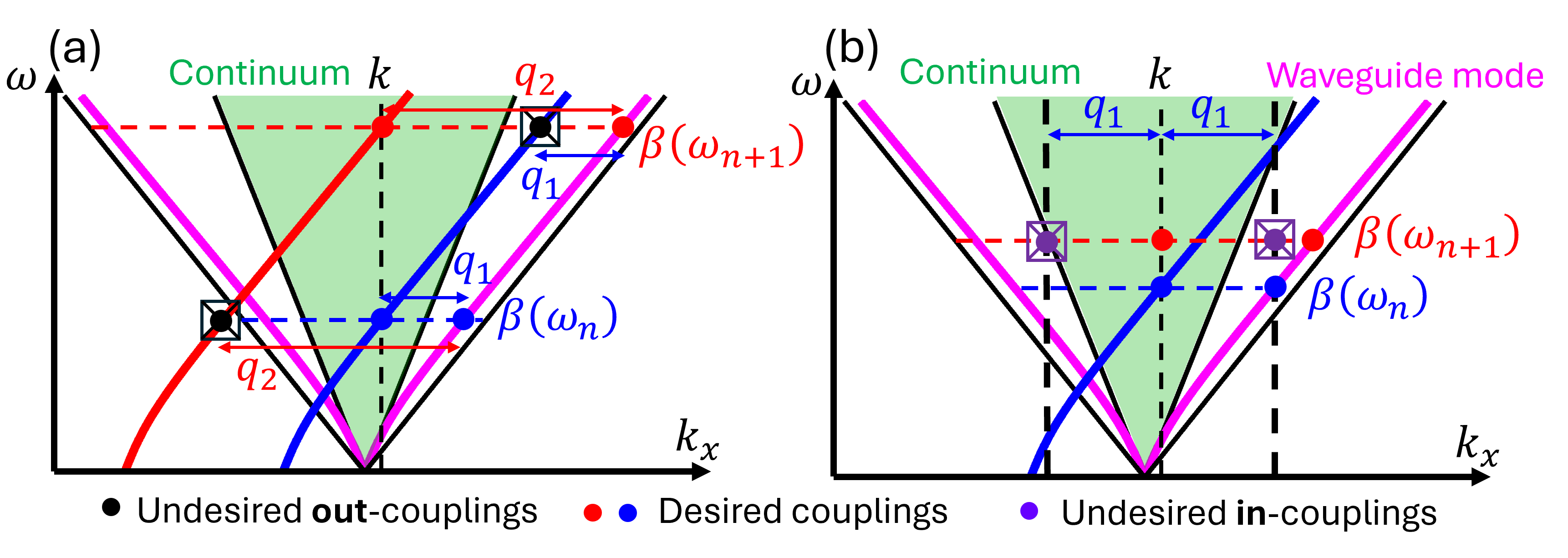}
    \caption{Band-diagrams illustrating the inequalities (a) Eq.~\ref{eq:sm_fig3_ineq1} prohibiting undesired \textit{out}-couplings and (b) Eq.~\ref{eq:sm_fig3_ineq2}} prohibiting undesired \textit{in}-couplings.
    \label{fig:sm_fig_1}
\end{figure}

For the undesired \textit{out}-couplings to lay outside the continuum, the following inequalities must be satisfied for any two successive frequencies $\omega_n,\omega_{n+1}$:
\begin{equation}
\begin{cases}
    \beta(\omega_{n+1})-q_n\geq \omega_{n+1}\\
    \beta(\omega_n)-q_{n+1}\leq-\omega_n 
\end{cases}
\overset{q_n=\beta(\omega_n)-k}{\iff}
\begin{cases}
    \beta(\omega_{n+1})-\beta(\omega_n)\geq\omega_{n+1}-k\\
    \beta(\omega_{n+1})-\beta(\omega_n)\geq \omega_n+k 
\end{cases},
\label{eq:sm_fig3_ineq1}
\end{equation}
where the first inequality prohibits the undesired out-coupling at frequency $\omega_{n+1}$ and the second inequality does the same for $\omega_n$, see Fig.~\ref{fig:sm_fig_1}. Assuming the frequency $\omega_n$ is fixed, there exists $\omega_{n+1}^\text{min}$ such that the inequalities Eq.~\ref{eq:sm_fig3_ineq1} are satisfied for any $\omega_{n+1}\geq\omega_{n+1}^\text{min}$. The latter minimal higher frequency $\omega_{n+1}^\text{min}$ is a function only of the lower frequency $\omega_n$, which we will denote by $\omega^\text{min}(\omega)$, which can be found from solving Eq.~\ref{eq:sm_fig3_ineq1} as equalities:
\begin{equation}
    \omega^\text{min}(\omega)=\text{max}(\omega_1,\omega_2),
    \text{ where }\omega_1,\omega_2\text{ are defined by }
    \begin{cases}
        \beta(\omega_1)-\beta(\omega)=\omega_1-k\\
        \beta(\omega_2)-\beta(\omega)=\omega+k
    \end{cases}
    \label{eq:sm_fig3_eq1}
\end{equation}
For the undesired \textit{in}-couplings to lay outside the continuum, we need
\begin{equation}
    \omega_n\leq \beta(\omega_1),\text{ for all } n.
    \label{eq:sm_fig3_ineq2}
\end{equation}
It is obviously sufficient to apply the inequality Eq.~\ref{eq:sm_fig3_ineq2} only to the highest frequency.

\begin{figure}[h]
    \centering
    \includegraphics[width=\linewidth]{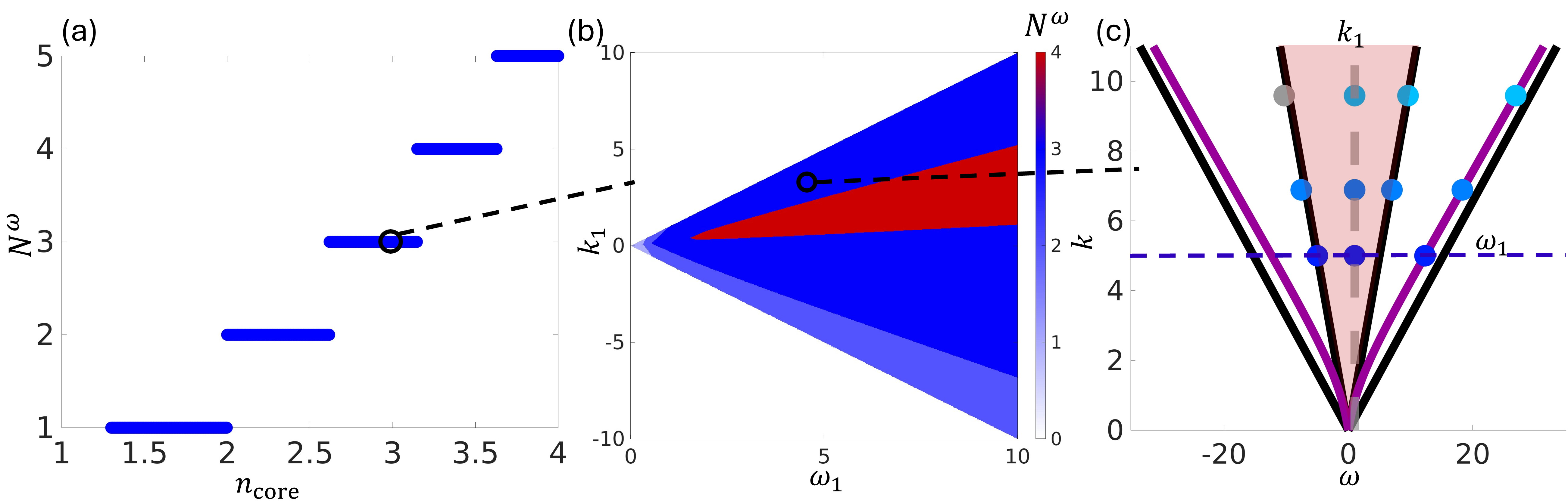}
    \caption{For every point in the plot (a) shown in Fig. 3 of the main text, a heatmap (b) of $N^\omega(\omega_1,k_1)$ was generated according to the procedure described above. (c) For every point in the heatmap, an optimal set of frequencies $\omega_1,...,\omega_n$ was generated according to the discussion above. Throughout the figure, $\omega,k$ are measured in units of $\frac{2\pi}{t^\text{wg}}$, where $t^\text{wg}$ is the total thickness of the grating coupler.}
    \label{fig:sm_fig1_2}
\end{figure}

To determine the maximum number of frequencies $\omega_1,...,\omega_{N^\omega}$ which can all simultaneously satisfy the inequalities of Eqs.~\ref{eq:sm_fig3_ineq1} and~\ref{eq:sm_fig3_ineq2}, we first choose some value for the lowest frequency $\omega_1$ and target wavenumber $k=k_1$, and by solving equations~\ref{eq:sm_fig3_eq1}, we find the next lowest $\omega_2$ satisfying inequalities~\ref{eq:sm_fig3_ineq1}. We then proceed by induction until at some step $n$ the next lowest frequency $\omega_n$ exceeds $\beta(\omega_1)$, violating inequality~\ref{eq:sm_fig3_ineq2}, at which point the process is terminated and the resulting $n$ is the desired maximum number of frequencies $N^\omega(\omega_1,k_1)$ for the selected initial guesses $\omega_1,k_1$. Finally, we repeat the described procedure for all (discretized) pairs of initial values $\omega_1\in[0,\omega^\text{max}]$ and $k_1\in[-\omega_1,\omega_1]$ for some very large cutoff frequency $\omega^\text{max}$ (e.g., $\omega^\text{max}/c=300\times \frac{2\pi}{d}$) the maximum:
\begin{equation}
    N^\omega=\max_{\omega_1\in[0,\omega^\text{max}],\;k_1\in[-\omega_1,\omega_1]} N^\omega(\omega_1,k_1).
\end{equation}
Carrying out the procedure described above for $n_\text{core}=3.0$ results in the heatmap for $N^\omega(\omega_1,k_1)$ shown in Fig.~\ref{fig:sm_fig1_2}, from which we can determine that $N^\omega|_{n_\text{core}=3}=4$. Similar data of $N^\omega(\omega_1,k_1)$ was generated for every value of $n_\text{core}$ shown in the main text.


\subsection{Parameter and constraint counting for multilayer couplers}

\begin{figure}[h]
    \centering
    \includegraphics[width=0.6\linewidth]{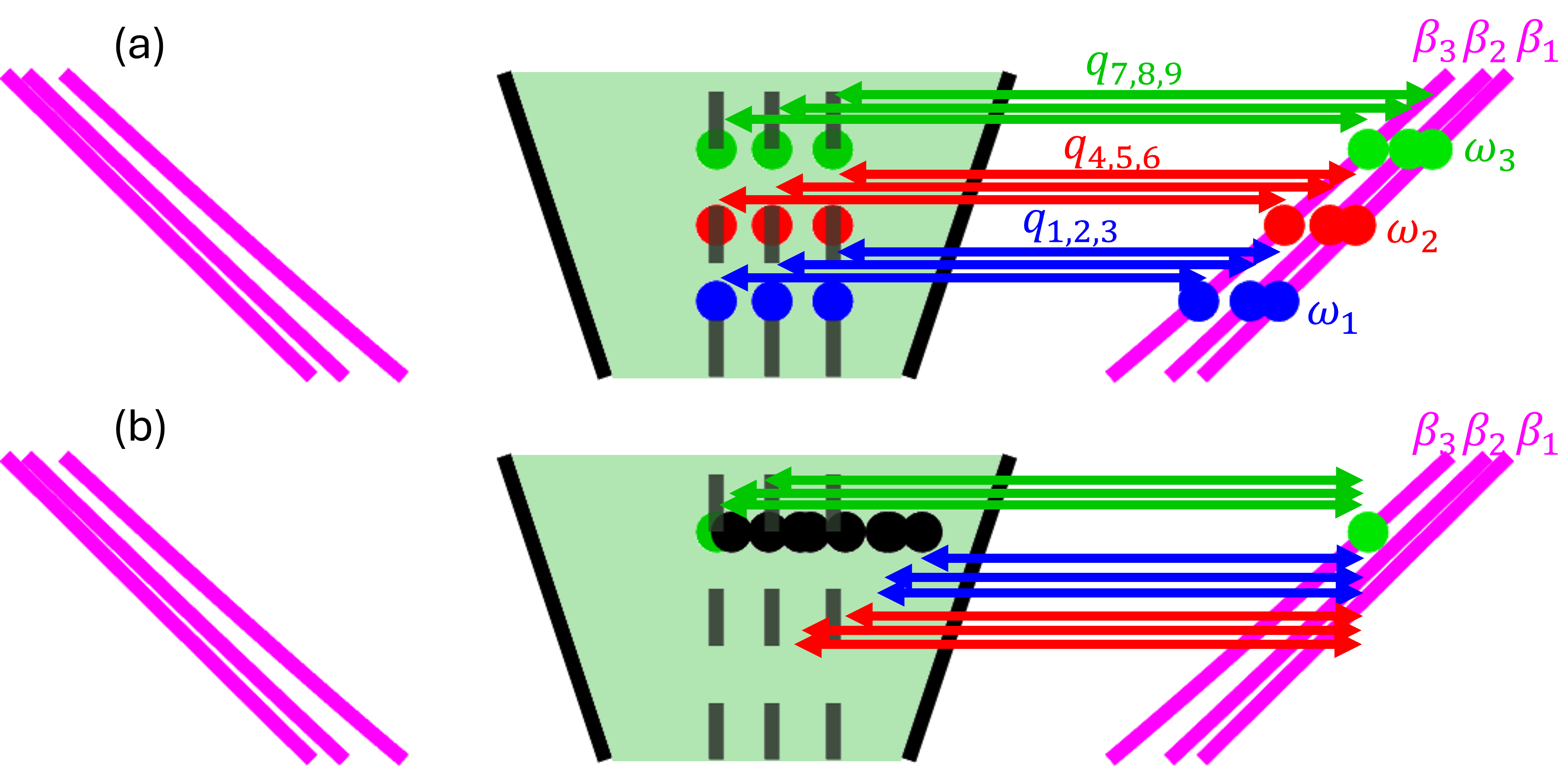}
    \caption{(a) Coupling $N^\text{fun}$ generic modes requires introducing $N^\text{fun}$ couplings. (b) $N^\text{fun}$ generic couplings result in $(N^\text{fun}-1)$ undesired couplings per desired coupling}
    \label{fig:sm_fig2}
\end{figure}

To couple $N^\text{fun}$ ``generic" modes, we need to introduce $N^\text{fun}$ Fourier components, since for frequencies and wavevectors in generic position we will have $q_{ij}=\beta_i(\omega_j)-k_i\neq\beta_n(\omega_m)-k_n=q_{nm}$ for any $(i,j)\neq(n,m)$. An example of such generic coupling is shown in Fig.~\ref{fig:sm_fig2} (a), for three frequencies and three wavevectors. Notice that above we temporarily deviated from the single-index notation $\vb q_n$ of the main text to emphasize the general many-frequency, many-wavevector coupling scenario to which the current considerations apply.

According to the idea of multilayer designs introduced in Section II of the main text, all the layers must share the same Fourier components $\vb q_i$, but the complex Fourier coefficients $C_i^{\text{layer}}$ are allowed to vary for each layer. Hence, for each layer we have $N^\text{fun}$ degrees of freedom and the total number of degrees of freedom is therefore
\begin{equation}
    N^\text{dof}=N^\text{fun}\times N^\text{layers}.
\end{equation}

Having introduced $N^\text{fun}$ Fourier components, we will create many undesired couplings in accordance with the ideas described in Section II of the main text. Generically, each Fourier component will enable one coupling from each waveguide mode, with only one of these being desired and the other $(N^\text{fun}-1)$ being undesired couplings, see Fig.~\ref{fig:sm_fig2} (b), resulting in a total of $N^\text{undes}=N^\text{fun}\times(N^\text{fun}-1)$ undesired couplings. The total number of coupling rates to be controlled in a multifunctional multilayer design is:
\begin{equation}
    N^\text{constraints}=2\left[N^\text{des}+N^\text{undes}\right]=2(N^\text{fun})^2,
    \label{eq:sm_eq_Ncon}
\end{equation}
where the factor of 2 takes account of the fact that both upward-going and downward-going couplings must be controlled (in high-efficiency couplers, waveguide modes must only be strongly scattering up, while downward-going radiation must be suppressed), and where $N^\text{undes}+N^\text{undes}$ is the total number of ``single-sided" couplings to be controlled, since undesired couplings must be suppressed while the desired coupling rates must be kept high.

\section{Supplementary Information to Section III}

\subsection{Limit on the number of couplings in single-layer 3D couplers}

\begin{figure}[h]
    \centering
    \includegraphics[width=0.5\linewidth]{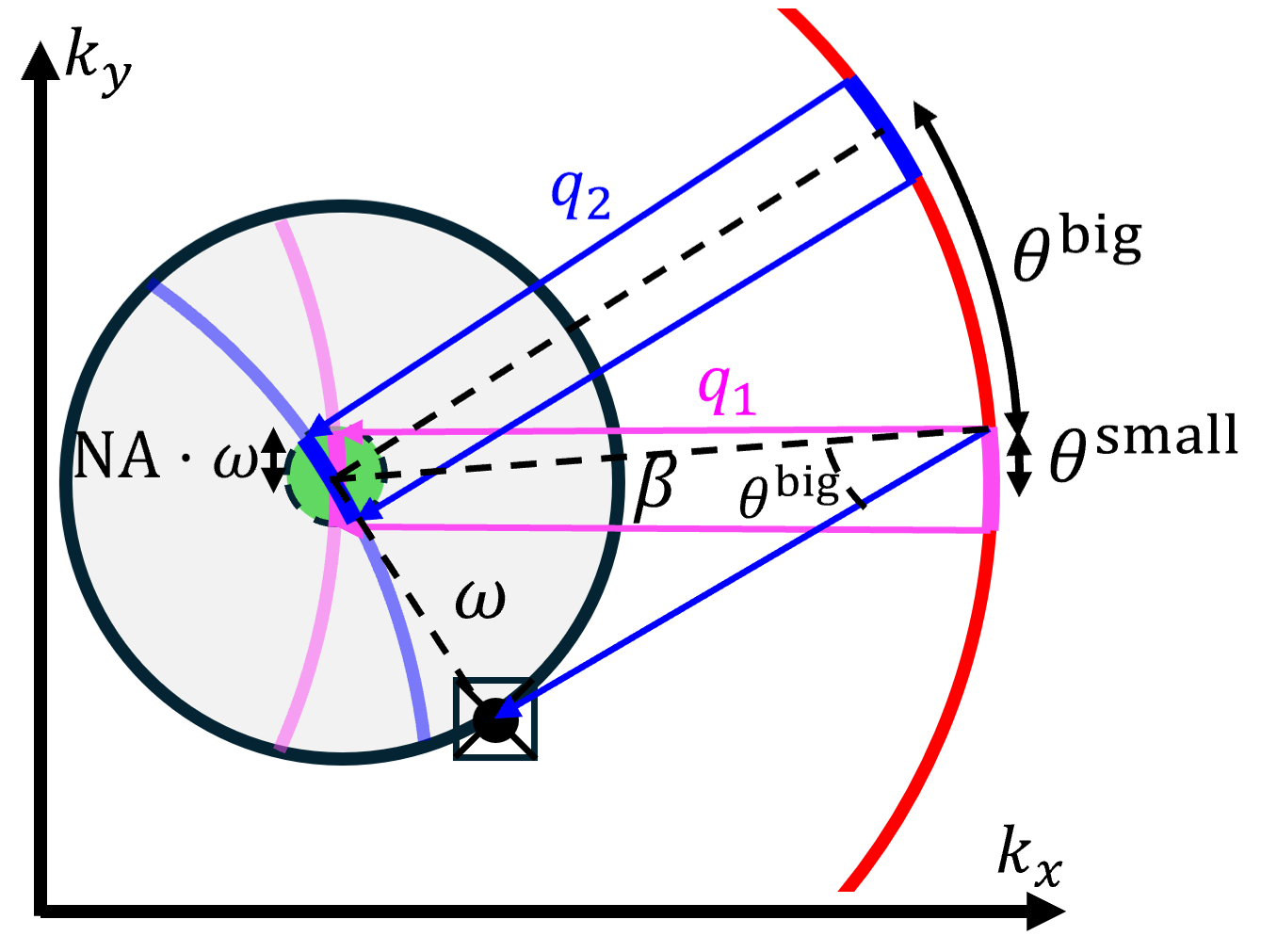}
    \caption{To avoid undesired couplings, $\vb q_2$ must be chosen such that every waveguide mode coupled via $\vb q_1$ to the desired numerical aperture is coupled via $\vb q_2$ to outside the continuum, resulting in a minimum required angle of $\Delta\theta^\text{min}=\theta^\text{big}+\theta^\text{small}$, with $\theta^\text{big}$ and $\theta^\text{small}$ shown in the figure.}
    \label{fig:sm_fig3}
\end{figure}

As described in Section III of the main text, to avoid undesired couplings, the angle between every pair of consecutive Fourier components $\vb q_n,\vb q_{n+1}$ of the 3D coupler must be large enough. The relevant geometry is shown in Fig.~\ref{fig:sm_fig3}: $\hat q_2$ must be chosen such that the line passing through the edge of the segment coupled by $\vb q_1$ (pink segment of the big circle) in the direction of $\hat q_2$ is tangential to the light cone (solid black circle). As a result, the minimum angle between two consecutive segments of waveguide modes which can be used for coupling is:
\begin{equation}
    \Delta\theta^\text{min}=\theta^\text{big}+\theta^\text{small},
    \label{eq:sm_sectII_A1}
\end{equation}
where
\begin{equation}
    \theta^\text{big}=\asin\frac{\omega}{\beta}=\asin\frac{1}{n^\text{eff}_\text{wg}}
    \label{eq:sm_sectII_A2}
\end{equation}
and
\begin{equation}
    \theta^\text{small}=\asin\frac{\omega\cdot\operatorname{NA}}{\beta}=\asin\frac{\operatorname{NA}}{n^\text{eff}_\text{wg}}
    \label{eq:sm_sectII_A3}
\end{equation}
are the angles shown in~\ref{fig:sm_fig3}. Note that $c=1$ was assumed throughout and $n^\text{eff}_\text{wg}=\frac{\beta}{\omega}$ is effective refractive index of the waveguide mode in question (e.g., fundamental TE mode for the coupling scenario considered in Section III of the main text).

Because the grating coupler design permittivity must be real, if the permittivity profile contains a Fourier component $\vb q$, it must also contain the Fourier component $-\vb q$, with possibly different Fourier coefficients $C_{\vb q},C_{-\vb q}$. Therefore, the maximum number of Fourier components that can be introduced without undesired couplings is given by twice the maximum number of integer multiples of $\Delta\theta^\text{min}$ that can fit in the $[0,\pi]$ interval (as opposed to the $[0,2\pi]$ interval which would have to be used without the $\pm\vb q$ symmetry). From Eqs.~\ref{eq:sm_sectII_A1}-\ref{eq:sm_sectII_A3}, the number of Fourier components $N^{\vb q}$ is thus
\begin{equation}
    N^{\vb q}=2\left\lfloor\frac{\pi}{\Delta\theta^\text{min}}\right\rfloor=2\left\lfloor\frac{\pi}{\asin\frac{1}{n^\text{eff}_\text{wg}}+\asin\frac{\operatorname{NA}}{n^\text{eff}_\text{wg}}}\right\rfloor
\end{equation}

\subsection{Number of modes coupled by 3D designs}

\begin{figure}[h]
    \centering
    \includegraphics[width=0.3\linewidth]{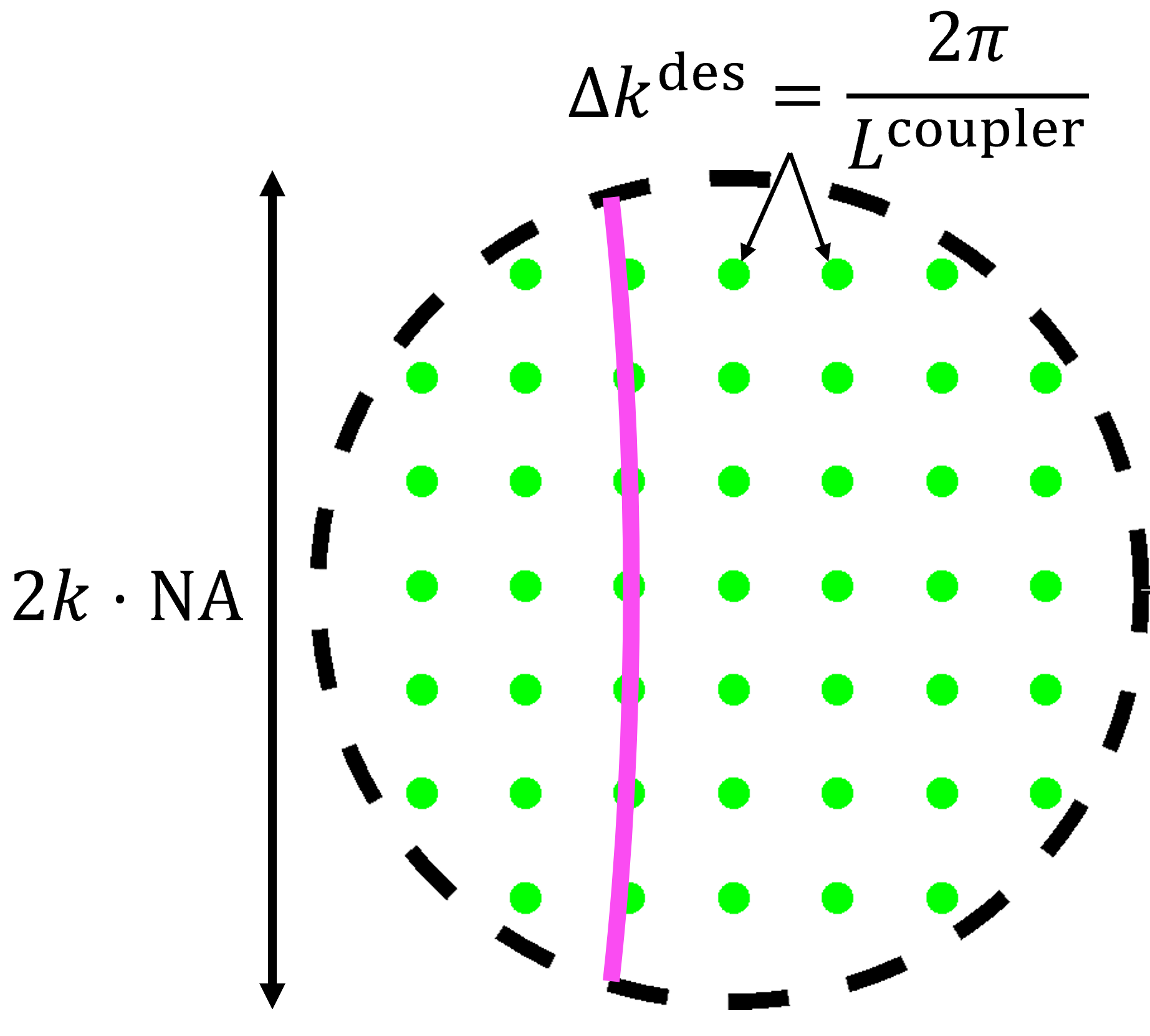}
    \caption{Spacing in momentum space between linearly independent free-space modes supported on the grating coupler is determined by the grating coupler size as $\frac{2\pi}{L^\text{coupled}}$. Total number of desired modes within a numerical aperture $\operatorname{NA}$ coupled by a single Fourier component is then given by Eq.~\ref{eq:sm_eq_Ntrivial}}
    \label{fig:sm_fig4}
\end{figure}

The number of modes coupled by a single coupling line shown in pink in Fig.~\ref{fig:sm_fig4} and in Fig. 5 of the main text can be determined from the spacing between desired modes
\begin{equation}
    \Delta k^\text{des}=\frac{2\pi}{L^\text{coupler}}
\end{equation}
and the length of the line in $k$-space, given by $2k\cdot\operatorname{NA}$, see Fig.~\ref{fig:sm_fig4}. The resulting number, denoted by $N^\text{trivial}$ in the main text is:
\begin{equation}
    N^\text{trivial}\propto \frac{2k\operatorname{NA}}{2\pi/L^\text{coupler}}=\frac{kL^\text{coupler}\operatorname{NA}}{\pi}\propto kL^\text{coupler}\operatorname{NA}.
    \label{eq:sm_eq_Ntrivial}
\end{equation}

The total number of modes coupled by $N^{\vb q}$ couplings is:
\begin{equation}
    N^\text{modes}\propto N^{\vb q}\times N^\text{trivial}\propto N^{\vb q}\cdot kL^\text{coupler}\cdot \operatorname{NA}.
\end{equation}

\subsection{Details of perturbative calculations for multifunctional 3D grating couplers}

\begin{figure}[h]
    \centering
    \includegraphics[width=0.7\linewidth]{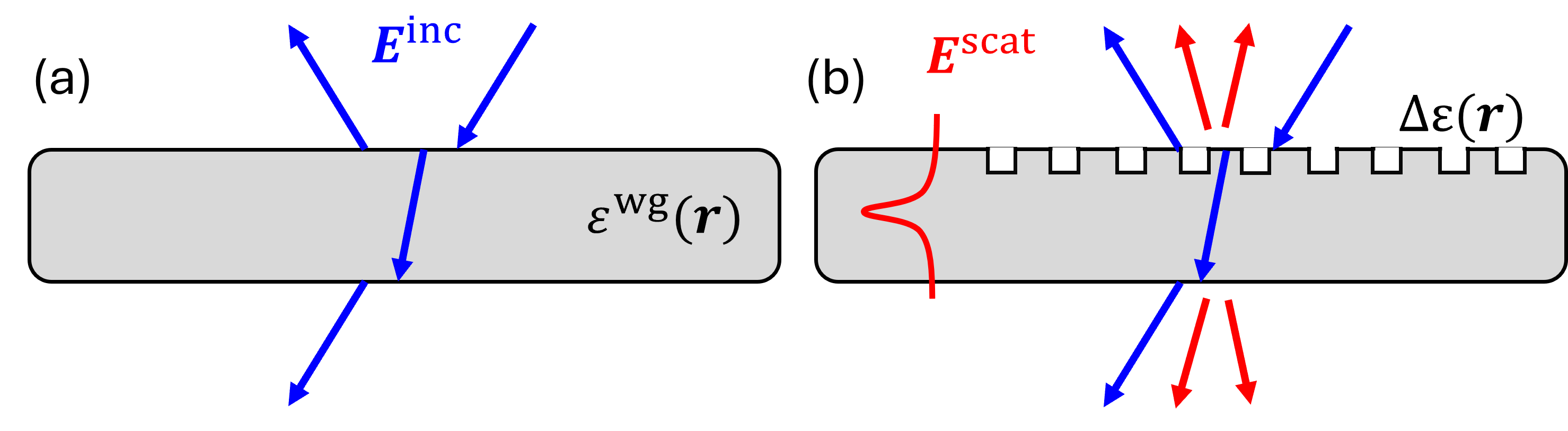}
    \caption{(a) Incident field is a solution to Maxwell's equations for the permittivity of an unpatterned slab waveguide $\epsilon^\text{wg}(\vb r)$. (b) The scattered fields result from the grating coupler patterning profile $\Delta\epsilon(\vb r)$}
    \label{fig:sm_fig5}
\end{figure}

In this subsection, we derive Eq. (8) of the main text - an expression for the scattering matrix of an arbitrarily patterned grating coupler shown in Fig.~\ref{fig:sm_fig5} in the limit when the patterning can be considered weak such that first-order perturbation theory may be applied. To do this, we write the wave equation for the electric field $\vb E(\vb r)$ in the frequency domain using the scattered field formulation~\cite{jianming}:
\begin{equation}
    [\curl\curl+k^2\epsilon^\text{gc}(\vb r)]\vb E^\text{scat}(\vb r)=-k^2\Delta\epsilon(\vb r)\vb E^\text{inc}(\vb r),
    \label{eq:sm_waveeq}
\end{equation}
where $k=\frac{\omega}{c}$ is the free-space wavenumber, $c=1$ will be assumed throughout the derivation; $\vb E^\text{inc}(\vb r)$ is the incident field - a solution to the Maxwell's equations in the presence of unpatterned waveguide, as shown in Fig.~\ref{fig:sm_fig5}; $\epsilon(\vb r)$ is the permittivity distribution of the grating coupler, and $\Delta\epsilon(\vb r)=\epsilon^\text{gc}(\vb r)-\epsilon^\text{wg}(\vb r)$ is the permittivity change introduced by the grating coupler pattern ($\epsilon^\text{wg}(\vb r)$ is the permittivity distribution of the unpatterned waveguide); finally, $\vb E^\text{scat}(\vb r)$ is the scattered field, satisfying the Silver-Muller radiation conditions at $|\vb r|\to\infty$~\cite{colton_kress}. The solution to Eq.~\ref{eq:sm_waveeq} may be written in terms of the dyadic electric and magnetic Green's functions of the grating coupler as~\cite{jianming}:
\begin{equation}
\begin{cases}
    \vb E^\text{scat}(\vb r)
    =
    -k^2\int \vb G^{\vb J_{\vb E}\to\vb E}_\text{gc}(\vb r,\vb r')\Delta\epsilon(\vb r')\vb E^\text{inc}(\vb r')d\vb r'\\
    \vb H^\text{scat}(\vb r)
    =
    -k^2\int \vb G^{\vb J_{\vb E}\to\vb H}_\text{gc}(\vb r,\vb r')\Delta\epsilon(\vb r')\vb E^\text{inc}(\vb r')d\vb r'
\end{cases}
    \label{eq:sm_Escat}
\end{equation}
where
\begin{equation}
\begin{cases}
    \vb G^{\vb J_{\vb E}\to \vb E}_\text{gc}=\left[\overset{\leftrightarrow}{I}+\frac{1}{k^2}\nabla\nabla\right]G_{gc}\\
    \vb G^{\vb J_{\vb E}\to \vb E}_\text{gc}=\frac{1}{ik}(\curl\overset{\leftrightarrow}{I})G_{gc}
\end{cases}
\end{equation}
and where $G_{gc}$ is the scalar Green's function of the grating coupler, satisfying
\begin{equation}
    [\nabla^2+k^2\epsilon^\text{gc}(\vb r)]G_\text{gc}(\vb r,\vb r')=\delta(\vb r-\vb r').
\end{equation}

Notice that the grating coupler Green's function $G_\text{gc}$ itself depends on $\Delta\epsilon(\vb r)$ in the sense that it may be expressed as~\cite{schotland_carminati}:
\begin{equation}
    G_\text{gc}=\left(I+k^2G_\text{wg}\Delta\epsilon\right)^{-1}G_\text{wg},
    \label{eq:sm_Ggc}
\end{equation}
which is nothing but the summed Born multiple scattering series.

To simplify the solution Eq.~\ref{eq:sm_Escat}, we expand it to first order in $\Delta\epsilon$, which requires taking the zeroth-order approximation $G_\text{gc}=G_\text{wg}$ in Eq.~\ref{eq:sm_Ggc}, and results in the following expressions for the fields:
\begin{equation}
    \begin{cases}
        \vb E_1^\text{scat}(\vb r)=-k^2\int \vb G^{\vb J_{\vb E}\to\vb E}_\text{wg}(\vb r,\vb r')\Delta\epsilon(\vb r')\vb E^\text{inc}(\vb r')d\vb r'\\
        \vb H_1^\text{scat}(\vb r)=-k^2\int \vb G^{\vb J_{\vb E}\to\vb H}_\text{wg}(\vb r,\vb r')\Delta\epsilon(\vb r')\vb E^\text{inc}(\vb r')d\vb r'
    \end{cases}
    \label{eq:sm_Escat_pert}
\end{equation}

\begin{figure}[h]
    \centering
    \includegraphics[width=\linewidth]{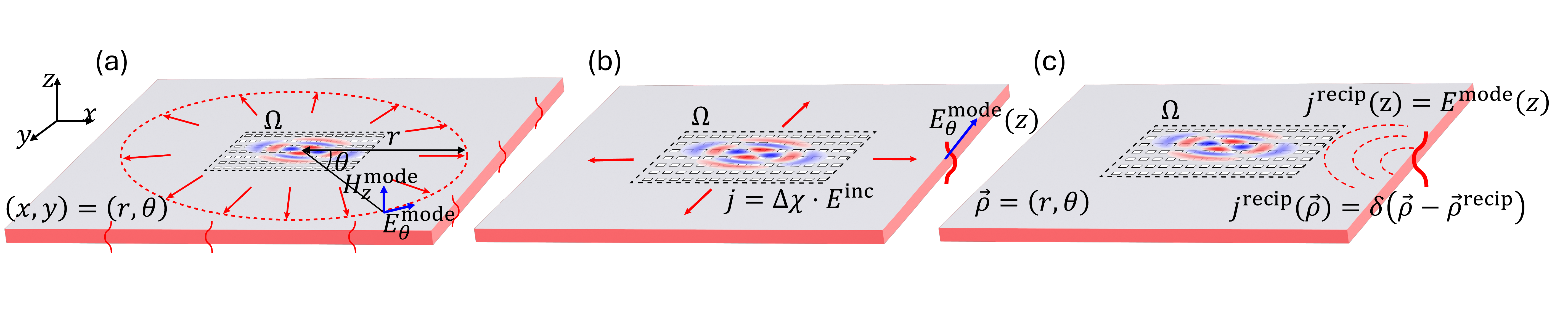}
    \caption{To evaluate (a) the coupling coefficient $c(r,\theta)$ of the scattered field with a slab waveguide mode defined on a line $\mathfrak{L}$ parallel to the $z$-axis going through $(r,\theta)$, we use Rayleigh-Carson reciprocity argument (b-c) to replace the integral of the scattered field over $\mathfrak{L}$ with an integral of the incident field over the grating coupler region $\Omega$.}
    \label{fig:sm_fig8}
\end{figure}

We define the coupling amplitude to a slab waveguide mode (defined on a line $\mathfrak{L}$ parallel to the $z$-axis) a distance $r$ away from the center of the grating coupler, in the azimuthal direction $\theta$, see Fig.~\ref{fig:sm_fig8} (a) as:
\begin{equation}
    c(r,\theta)=
    \sqrt{r}\left.
    \bra{
    \begin{pmatrix}
    \vb E^\text{mode}\\
    \vb H^\text{mode}
    \end{pmatrix}
    }
    \ket{
    \begin{pmatrix}
        \vb E^\text{scat}\\
        \vb H^\text{scat}
    \end{pmatrix}
    }
    \right|_{\mathfrak{L}}
    =
    \sqrt{r}\frac{1}{4}\int_{-\infty}^{+\infty}
    \begin{pmatrix}
    \vb E^\text{mode} \\ \vb H^\text{mode}
    \end{pmatrix}^\dag
    \begin{pmatrix}
        0 & -\hat r\cross\\
        \hat r\cross & 0
    \end{pmatrix}
    \begin{pmatrix}
    \vb E^\text{scat} \\ \vb H^\text{scat}
    \end{pmatrix}
    dz,
    \label{eq:sm_cdef}
\end{equation}
where
\begin{equation}
\begin{cases}
    \vb E^\text{mode}(z)=E^{\text{TE}_0}(z)\cdot \hat \theta\\
    \vb H^\text{mode}(z)
    =
    H^{\text{TE}_0}_\perp(z)\cdot \hat z
    +
    H^{\text{TE}_0}_\parallel(z)\cdot \hat r
\end{cases}
\label{eq:sm_EHmode}
\end{equation}
are the fundamental TE waveguide mode profiles, power-normalized such that $\frac{1}{2}\int E^{\text{TE}_0}(z)H_\perp^{\text{TE}_0}dz=1$. We adopt the definition Eq.~\ref{eq:sm_cdef} because the overall coupling efficiency of the incident mode $\vb E^\text{inc}$ to the slab waveguide can then be expressed as
\begin{equation}
    \operatorname{CE}=\int_0^{2\pi} |c(r,\theta)|^2d\theta.
\end{equation}

Exchanging the order of integration with respect to $\vb r$ and $\vb r'$ in Eq.~\ref{eq:sm_cdef} and using a standard Rayleigh-Carson reciprocity argument~\cite{jianming}, we can re-write the coupling coefficient $c(r,\theta)$ as an integral over the grating coupler region. To set up the reciprocity argument, we write:
\begin{equation}
    c(r,\theta)\frac{1}{\sqrt{r}}=\int_{\mathfrak{L}} d\vb r \vb j^{\text{mode}_{r,\theta}}(\vb r)\cdot\int_\Omega \vb G_\text{gc}(\vb r,\vb r')\vb j^\text{inc}(\vb r')d\vb r',
    \label{eq:sm_randomeq1}
\end{equation}
where $\Omega$ is the grating coupler region and $\mathfrak{L}$ is the line on which the considered waveguide mode is defined, where the ``incident'' electric and magnetic sources $\vb j^\text{inc}(\vb r')$ are defined using Eq.~\ref{eq:sm_Escat_pert} as:
\begin{equation}
    \vb j^\text{inc}_{\vb E}(\vb r')=-k^2\Delta\epsilon(\vb r')\vb E^\text{inc}(\vb r'),\;\vb j^\text{inc}_{\vb H}(\vb r')=0,\;\vb r'\in\Omega,
\end{equation}
and where the ``modal'' sources are defined as:
\begin{equation}
\begin{cases}
    \vb j^{\text{mode}_{r,\theta}}_{\vb E}(r',\theta',z)=-\hat r\cross(\vb H^\text{mode}(z))^*\delta(r-r',\theta-\theta')\\
    \vb j^{\text{mode}_{r,\theta}}_{\vb H}(r',\theta',z)=\hat r\cross(\vb E^\text{mode}(z))^*\delta(r-r',\theta-\theta')
\end{cases}
\label{eq:sm_modalsources}
\end{equation}
Exchanging the order of integration and using Rayleigh-Carson theorem in Eq.~\ref{eq:sm_randomeq1}, we obtain:
\begin{equation}
    c(r,\theta)\frac{1}{\sqrt{r}}=-k^2\int_\Omega \vb E^{\text{mode}_{r,\theta}}\cdot\Delta\epsilon(\vb r')\vb E^\text{inc}(\vb r')d\vb r'
    \label{eq:sm_randomeq2}
\end{equation}
where the field $\vb E^{\text{mode}_{r,\theta}}$ is the electric field generated by the sources Eq.~\ref{eq:sm_modalsources} in an unperturbed waveguide. Notice that the modal sources Eq.~\ref{eq:sm_modalsources} are written in the form of the surface equivalence principle~\cite{jianming} and if these sources were distributed across a surface with normal $\hat r$, the resulting field would simply be $(\vb E^\text{mode})^*$. However, because the sources are distributed across the line $\mathfrak{L}$, the field $\vb E^{\text{mode}_{r,\theta}}$ decomposes into the transverse TE fundamental mode profile and a 2D profile due to a dipole source at $(r,\theta)$:
\begin{equation}
    \vb E^{\text{mode}_{r,\theta}}(r',\theta',z)=E^{\text{TE}_0}(z)H_0^{(1)}(\beta\norm{(r,\theta)-(r',\theta')}),
\end{equation}
where $\beta$ is the propagation constant of the fundamental TE mode of the slab waveguide. Taking the point $(r,\theta)$ where we are trying to evaluate the coupling amplitude $c(r,\theta)$ to be in the far field, we have $\beta \norm{(r,\theta)-(r',\theta')}\gg 1$ and using the asymptotic formulas for the Hankel functions, we have
\begin{equation}
    \vb E^{\text{mode}_{r,\theta}}(r',\theta')=\sqrt{\frac{2}{\beta r}}e^{i\frac{\pi}{4}}e^{-i\beta\hat r\cdot\vb r'},.
\end{equation}
Substituting this back into Eq.~\ref{eq:sm_randomeq2}, we have
\begin{equation}
    c(r,\theta)=\alpha\int_\Omega e^{-i\beta\hat r\cdot\vb r'}\Delta\epsilon(\vb r')\vb E^\text{inc}(\vb r')d\vb r',\text{ where }\alpha=\sqrt{\frac{2}{\pi\beta}}e^{\frac{\pi}{4}}.
    \label{eq:sm_cfinal}
\end{equation}
We can cast this equation in the form of Eq. (8) of the main text by introducing matrices containing the waveguide and incident mode profile distributions. For this, let's discretize the grating coupler region such that $\vb r'\in \Omega$ is a discrete set and such that $\Delta\vb r'$ is the discretized volume element. Suppose incident fields are labeled by their incident wavevector $\vb k^\text{inc}$ and waveguide modes are labeled by the direction of their propagation $\hat r$. Then define matrices
\begin{equation}
    M^\text{wg}(\hat r,\vb r'):=\sqrt{\frac{2}{\pi\beta}}e^{-i\frac{\pi}{4}}e^{i\beta\hat r\cdot\vb r'}\text{ and }M^\text{free}(\vb k^\text{inc},\vb r')=\vb E^\text{inc}_{\vb k^\text{inc}}(\vb r'),
\end{equation}
and view $\Delta\epsilon(\vb r')\Delta\vb r'$ as a diagonal matrix $\chi(\vb r',\vb r''):=\Delta\epsilon(\vb r')\Delta\vb r'\cdot \delta(\vb r'-\vb r'')$, where $\delta(\vb r'-\vb r'')$ is the Dirac delta function. Finally, notice that $c(r,\theta)$ is nothing but the $(\hat r,\vb k^\text{inc})$-entry of the $S$-matrix of the grating coupler. Using our newly introduced matrix notation, we can write the integral for this matrix element in compact matrix notation: 
\begin{equation}
    S=(M^\text{wg})^\dag\chi M^\text{free}.
\end{equation}

For more details on how Fig. 6 of the main text was generated, see the corresponding code~\cite{github_code}. 

\subsection{Details of full-wave simulation of high-efficiency multi-functional couplers}

The permittivity of the design considered in Fig. 7 of the main text is given by:
\begin{equation}
    \epsilon(\vb r)=1+(\epsilon^\text{max}-1)\cdot \frac{1+\frac{1}{4}\sum_{n=1}^4\cos(\vb q_n\cdot\vb r)}{2},
    \label{eq:sm_eps}
\end{equation}

\begin{figure}[h]
    \centering
    \includegraphics[width=0.9\linewidth]{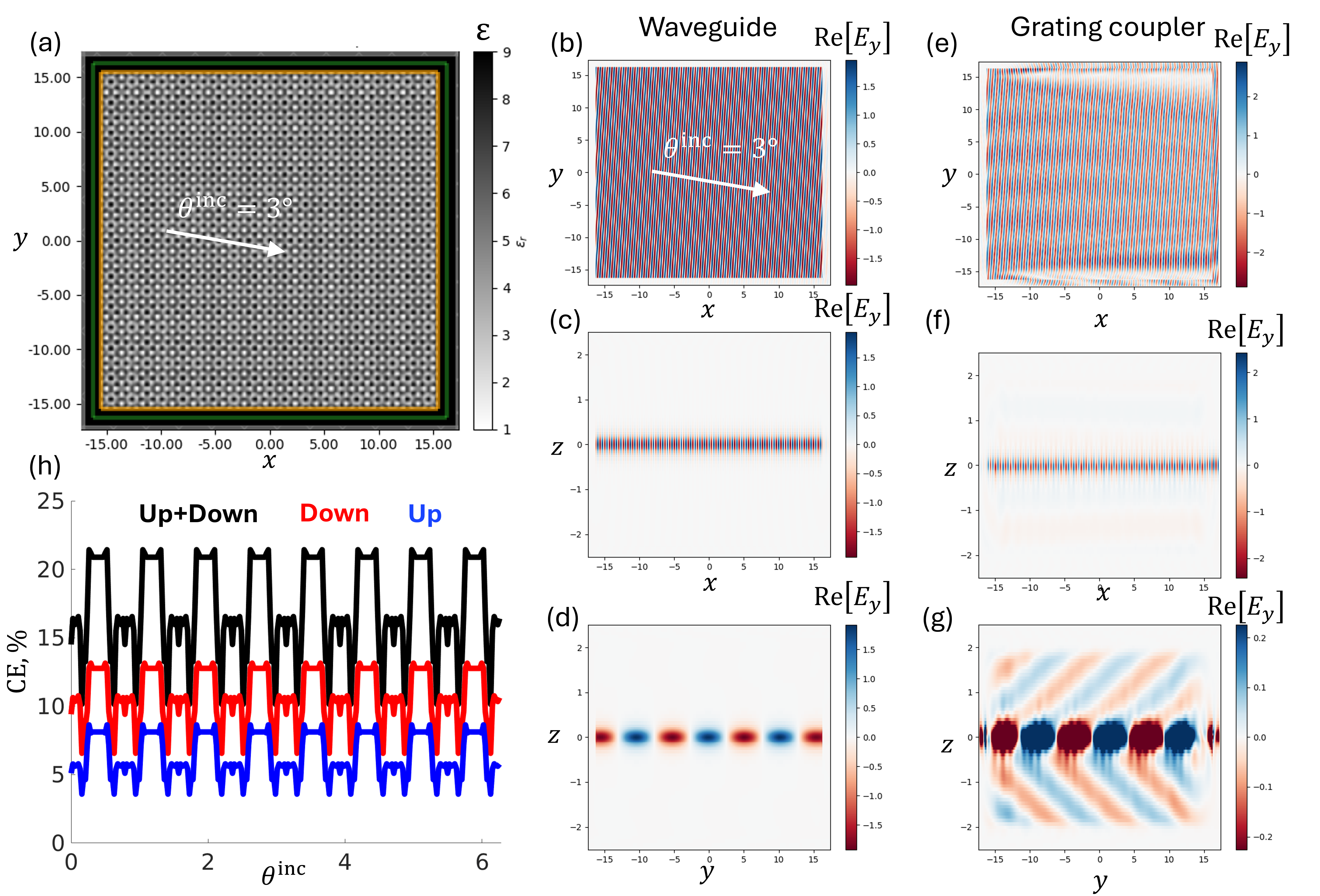}
    \caption{(a) Design permittivity given by Eq.~\ref{eq:sm_eps}. (b-d) Fields of the incident slab waveguide mode in an unpatterned waveguide, created using an array of modal sources on the boundary of the grating coupler region. (e-g) The total fields due to the same sources as in (b-d) in the presence of the grating coupler. (h) Coupling efficiencies up and down calculated using field monitors above and below the grating coupler. If the grating coupler is patterned on top of the waveguide, the coupling efficiency down turns out to be higher than up.}
    \label{fig:sm_fig10}
\end{figure}

where $\epsilon^\text{max}=(n^\text{max})^2$ is the maximum permittivity corresponding to a maximum refractive index $n^\text{max}=3.0$ and where $\vb q_n=\beta\cdot (\cos\frac{\pi}{4}n,\sin\frac{\pi}{4}n)|,n=1,2,3,4$, where $\beta$ is the propagation constant of the fundamental mode of the slab with thickness $t^\text{wg}=0.22\;\mu m$ when the operating wavelength is $\lambda=1550\;\text{nm}$ . The slab waveguide mode was launched using an array of dipole sources on the two-dimensional boundary of the grating coupler (including the out-of-waveguide-plane direction) with dipole amplitudes varying according to the fundamental TE mode profiles, see Fig.~\ref{fig:sm_fig10}. Both upward- and downward-going radiation was calculated using field monitors, see Fig.~\ref{fig:sm_fig10}. It was found that if the grating coupler was patterned on the top of the waveguide, more radiation was scattered down than up, see Fig.~\ref{fig:sm_fig10}, therefore coupling down was chosen as the primary direction and coupling efficiencies reported in the main text are for coupling down. Notice that coupling up can be achieved by patterning the bottom of the waveguide instead.

To view the Tidy3D notebook used to generate this figure and Fig. 7 in the main text, see~\cite{github_code}.

\bibliography{SM_refs}